\documentclass[preprint,12pt]{elsarticle}
\journal{Nuclear Physics B}
\usepackage{amssymb,graphicx,amsmath,graphics,xfrac,url,booktabs,adjustbox,textcomp,amsfonts}
\usepackage{caption}
\usepackage{subcaption}
\usepackage{xcolor}

\newcommand\sca{\mathrm{S}}

\newcommand\ten{\mathrm{T}}

\begin{document}

\begin{frontmatter}

\title{Observational predictions of some inflationary models}

\author[label1]{Daniel Pozo}
\ead{daniel.pozo@yachaytech.edu.ec}
\affiliation[label1]{organization={Yachay TECH University,School of Physical Sciences and Nanotechnology},
            addressline={Hda. San Jose S/N y Proyecto Yachay}, 
            city={Urcuqui},
            postcode={100119}, 
            state={Imbabura},
           country={Ecuador}}
\fntext[label3]{}

\author[label2]{Lenin Calvache}

\author[label3]{Esteban Orozco}
\affiliation[label2]{organization={Instituto de Radioastronomia y Astrofisica, UNAM},
            addressline={Campus Morelia, AP 3-72}, 
            city={Mexico},
            postcode={58089},
           country={Mexico}}

\author[label4]{Vicente A. Ar\'evalo}

\author[label5]{Clara Rojas}

\begin{abstract}
This paper presents the CMB angular power spectrum obtained using the \texttt{CAMB} code for three different models of inflation: the Starobinsky inflationary model, the generalized Starobinsky inflationary model, and the chaotic inflationary model with a step. The results are compared with the most recent data reported for the Planck mission. An analysis of the large ($\ell \lesssim 90$), intermediate ($90 \lesssim \ell \lesssim 900$), and small ($\ell \gtrsim 900 $) angular scales is performed. We report the position of the peaks in the intermediate region so as the cosmological parameters obtained in each of the models: age of the universe, $\Omega_m$, $\Omega_b$, $\Omega_{\Lambda}$, $\Omega_K$ and $n_\sca$. We also perform a Bayesian analysis using the \texttt{Cobaya} code to evaluate our three best-fitting models. Additionally, we generated contour plots $(n_\sca,r)$ for our inflationary models, taking into account the number of e--folds between the end of inflation and the completion of reheating.

\end{abstract}

\begin{keyword}
CMB angular power spectrum \sep \texttt{CAMB} code \sep \texttt{Cobaya} code \sep Reheating \sep Starobinsky inflationary model \sep generalized Starobinsky inflationary model \sep chaotic inflationary model with a step.
\end{keyword}

\end{frontmatter}
\section{ Introduction}

The Big Bang theory describes the development of the Universe from a much hotter and denser point to the present. It is supported by solid observational evidence: the expansion of the Universe, the prediction of light element abundances, and the existence of the Cosmic Microwave Background (CMB) radiation. 
The  CMB is a picture of the recombination epoch ($370,000$ years after the big bang). The first prediction of the CMB was made by Alpher and Herman in $1948$ \cite{alpher:1948}, and was discovered in $1965$ by Penzias and Wilson \cite{penzias:1965}. It became one of the most important observational probes of the Big--bang theory. The CMB is an almost uniform and isotropic radiation field and shows a perfect black body spectrum at a temperature of $2.72$K \cite{tojeiro:2006}. The CMB has fluctuations in a part of $10^{5}$ that gives rise to the structure of the present Universe \cite{durrer:2015}.

Although Big Bang theory stands as a robust and widely accepted explanation for the beginning of our universe, this theory has three problems related to its initial conditions: the flatness problem, the horizon problem, and the monopole problem. To address these concerns, Alan Guth introduced the inflationary theory in $1981$ \cite{guth:1981}. This theory proposes an epoch of accelerated expansion preceding the hot stage mentioned in the traditional Big Bang model. The simplest way to produce such an epoch is via the potential energy of a scalar field, and the field $\phi$ is called the inflaton. There are several inflationary models with their own scalar potentials, and over the years studies have been made on which of them is favored by satellite data \cite{pozo:2023,akrami:2020,martin:2014a,martin:2006}. 
One notable outcome of inflation is that it provides a mechanism to generate the primordial scalar perturbations and produce an almost scale-invariant scalar power spectrum.

The accurate recreation of the angular power spectrum can judge how well an inflationary theoretical model works to describe our early Universe, in this work this recreation is made using the \texttt{CAMB} code \cite{Lewis:1999bs,Howlett:2012mh}. The recent report by the Planck mission favors the models with a low amount of tensor perturbations i.e., small values of tensor--to--scalar ratio\cite{aghanim:2020}. For this reason, the Starobinsky inflationary model \cite{starobinsky:1980} has become an important candidate, however a slight variation of the Starobinsky inflationary model that depends on a parameter $p$ close to the unity improved our results \cite{rojas:2022,meza:2021,renzi:2020}.  Recently, Di Valentino and Mersini--Houghton \cite{divalentino:2017} have studied the CMB angular power spectra for the modified Starobinsky potential in the context of the quantum landscape multiverse using a modified \texttt{CAMB} code. 

We also analyze the chaotic inflationary model with a step, which has been caused of interest in recent years \cite{thomas:2023,hernandez:2023,mishra:2020,cadavid:2015,adams:2001}. Adams and Cresswell \cite{adams:2001}, influenced by the existence of potentials that lead to scale dependent spectra, studied the consequences of introducing a step in the quadratic potential of the chaotic inflationary model, $V(\phi) = \frac{1}{2} m^2\phi^2$. Adding ``features'' to the potential can produce scale dependence in the primordial perturbation spectra \cite{adams:2001}.

Using the \texttt{CAMB} code we study for each model the three regions of the CMB angular power spectrum: a) the Sachs--Wolfe plateau region ($\ell < 90$) which correspond to large angular scales ($\theta > 2^\circ$), b) the acoustic peak region ($90 < \ell < 900$), and c) the silk damping region ($\ell > 900$).
It is important to note that Macorra \textit{et al.} have used the \texttt{CAMB} code with full \texttt{COSMO--MC} (\texttt{MCMC}) calculation to give a solution to one of the most recent conflicts in Modern Cosmology: the Hubble tension \cite{macorra:2022}. Also, Purkayastha et al. have estimate CMB E mode signal over large angular scales \cite{purkayastha:2022}. Additionally, we have used the \texttt{Cobaya} code \cite{torrado:2021} in order to perform a Bayesian analysis of our three best inflationary models.

To study CMB, we utilize the radiation angular power spectrum, which is measurable and theoretically predictable. The radiation angular power spectrum $C_\ell$ is defined by:

\begin{equation}
C_\ell=\left\{ \left|a_{\ell m}\right|^2\right\},
\end{equation}
where $a_{\ell m}$ are the coefficients of the expansion in spherical harmonics

\begin{equation}
\dfrac{\Delta T}{T}\left(\theta,\phi\right)=\displaystyle \sum_{\ell=1}^{\infty}\sum_{m=-\ell}^{\ell}a_{\ell m} Y_m^{\ell}(\theta,\phi),
\end{equation}
of a dimensionless temperature anisotropy, which is given by

\begin{equation}
\dfrac{\Delta T}{T}(\theta,\phi)=\dfrac{T(\theta,\phi)-{\overline{T}}}{\overline{T}},
\end{equation}
where $T(\theta,\phi)$ is the microwave background temperature measured in some direction on the sky and $\overline{T}$ the mean temperature.

$C_\ell$ depends only on $\ell$ since the statistical properties are required to be independent of the choice of the origin of $\theta-\phi$, that is, rotational invariance, or from a physical point of view, isotropy \cite{serjeant}. The larger is $\ell$, the smaller is the angular scale at which the spherical harmonics have variation \cite{liddle}.

Also (and this is the case for this article), the $C_\ell$ spectrum is conventionally plotted as $\ell(\ell+1)C_\ell\overline{T}^2/{2\pi}$, measuring the power per logarithmic interval in $\ell$. In such plot, a scale--invariant spectrum looks horizontal \cite{serjeant}.

Using the \texttt{CAMB} code we study, for each model, the three regions of the CMB angular power spectrum:

\subsection{The Sachs-–Wolfe plateau region ($\ell$ $<$ 90)}

At this scale (large scale), also called the horizon scale, anisotropies reflect the initial conditions since they have not evolved significantly.

Assuming a nearly scale--invariant spectrum of density perturbations in early times ($n_S \approx 1$), meaning gravitational potential fluctuations that are independent of $k$, then $\ell(\ell + 1)C_\ell$ remains constant at low $\ell_S$. This effect is more evident when the multipole axis is plotted logarithmically, as we did in this article.

Time variation in potentials, associated with time--dependent metric perturbations, causes an increase in the $C_{\ell S}$ at the lowest multipoles, influenced by any deviation from a total equation of state $w = 0$. The Dark Energy \textquotesingle s dominance at low redshift causes the lowest ls to rise above the plateau, known as the `integrated Sachs--Wolfe effect' or ISW Rise, confirmed through correlations between large--angle anisotropies and large--scale structure.

In summary, the mechanism generating primordial perturbations produces scalar, vector, and tensor modes. While vector modes and tensors decay with the expansion of the Universe, their contribution to the low $\ell$ signal, especially gravity waves, can be challenging to distinguish from other effects. However, polarization information can help to differentiate tensor modes \cite{scott}.

\subsection{The acoustic peak region ($90< \ell < 900)$}

The intricate patterns observed in the anisotropy spectrum on intermediate scales can be attributed to gravity--driven acoustic oscillations that occurred prior to the neutralization of atoms in the universe. The frozen--in phases of these sound waves leave an imprint on the cosmological parameters, lending significant constraining power to the anisotropies in the CMB.

During the era when the proton--electron plasma was closely coupled to photons, both components behaved as a unified `photon--baryon fluid', wherein photons contributed primarily to pressure and baryons imparted inertia. Gravitational potential perturbations, dominated by dark matter, drove oscillations in this fluid, with photon pressure acting as the restoring force. Despite their small amplitude ({$\mathcal{O}$}($10^{-5}$)), these perturbations evolved linearly, treating each Fourier mode independently as a driven harmonic oscillator. The frequency of these oscillations was determined by the sound speed in the fluid, resulting in a fluid density oscillation with a velocity out of phase and a reduced amplitude.

Following the recombination of baryons and the decoupling of radiation, allowing photons to travel freely, the oscillation phases were frozen--in. Projected onto the sky, these frozen--in phases manifested as a harmonic series of peaks. The primary peak represented the mode that completed $\sfrac{1}{4}$ of a period, reaching maximum compression. Even peaks denoted maximal under--densities, typically of smaller amplitude due to the rebound having to overcome baryon inertia. The troughs, which did not reach zero power, were partially filled as they corresponded to the maximum velocity \cite{scott}.

\subsection{The silk damping region  ($\ell > 900)$}

The last scattering surface acquires a non--instantaneous thickness due to the recombination process, resulting in a damping effect on anisotropies at higher l\textquotesingle s. This damping is particularly noticeable on scales smaller than the thickness of the last scattering surface. Another perspective is to view the photon--baryon fluid as having imperfect coupling, leading to diffusion between the two components and a gradual reduction in the amplitudes of oscillations over time. These effects collectively contribute to a phenomenon known as Silk Damping, effectively truncating the anisotropies at multipoles exceeding approximately 2000 \cite{scott}.

The paper is organized as follows: in Section 2 we define the slow--roll parameters and we describe how to calculate the cosmological parameters into the slow--roll approximation.
Sections $3$, $4$, and $5$ are devoted to calculating the CMB angular power spectrum obtained from \texttt{CAMB} for the Starobinsky inflationary model, the generalized Starobinsky inflationary model, and the chaotic inflationary model with a step. {In Section $6$ we made a comparison of our three best fitting models through the calculation of $\chi^2$. Section $7$ shows the results of the Bayesian Analysis with the variation of the effective number of neutrino species $\Delta N_\text{eff}$ in order to  determine which of our three best fitting models is most favored. Section 8 focuses on the incorporation of reheating analysis and its predictions in the $(n_\sca,r)$ contour plots for all of our inflationary models. Finally, the conclusions are discussed in Section $9$.

\section{Slow--roll approximation and cosmological parameters}

The slow--roll approximation is a standard technique in inflationary cosmology to solve the movement equations and the scalar and tensor equation of perturbations. This approximation is characterized by a set of parameters that establish when the approximation is valid or not and are given in terms of the inflationary potential and its derivatives in the following way \cite{lazaroui:2024,zarei:2016}:

\begin{eqnarray}
\label{epsilon}
\epsilon_\nu &=&\dfrac{1}{2}\left(\dfrac{V'}{V}\right)^2 ,\\
\eta_\nu &=& \dfrac{V''}{V},\\
\xi_\nu^2 &=& \dfrac{V' V'''}{V^2},\\
\label{omega}
\omega_\nu^3 &=&  \dfrac{V'^2 V''''}{V^3},
\end{eqnarray}
where $\epsilon_\nu$, $\eta_\nu$, $\xi_\nu^2$, and $\omega_\nu^3$ are known as the slow--roll parameters, while $V$ represents the potential characterizing the inflationary model under consideration. The prime denotes a derivative with respect to the scalar field $\phi$.

In order to compute the CMB power spectrum we need to configure each .ini file for the \texttt{CAMB} code, it is essential to determine the initial value of the scalar field at the beginning of inflation $\phi_\text{ini}$ \cite{zarei:2016}. This value can be computed using the expression for the number of e-folds, $N_e$, under the slow--roll approximation, as follows:

\begin{equation}
\label{e-folds}
N_e\simeq \int_{\phi_\text{end}}^{\phi} \dfrac{V}{V^\prime} \mathrm{d} \phi,
\end{equation}
where  $\phi_{\text{end}}$ is the value of the scalar field at the end of inflation, determined by the condition $\epsilon_\nu = 1$, where $\epsilon_\nu$ is one of the slow--roll parameters defined in Eq. \eqref{epsilon}. The number of e--folds, $N_e$, is fixed at 60 \cite{zarei:2016}.

Using the value of $\phi_\text{ini}$, we calculate the cosmological quantities  in terms of the slow--roll parameters as,

\begin{eqnarray}
\label{nS}
n_\sca &\simeq& 1 - 6 \epsilon + 2\eta,\\
\alpha_\sca &\simeq&
16\eta\epsilon-24\epsilon^2-2\xi^2,\\
\beta_\sca &\simeq&
192\epsilon^3-192\epsilon^2\eta+32\epsilon\eta^2+24\epsilon\xi^2-2\eta\xi^2-2\omega^3,\\
n_\ten &\simeq& -2\epsilon ,\\
\alpha_\ten &\simeq& 4\eta\epsilon-8\epsilon^2,\\
\label{r}
r &\simeq& 16\epsilon.
\end{eqnarray}

In addition to the initial conditions for the scalar field, we also require the amplitudes of the scalar power spectrum as input for the \texttt{CAMB} code. To perform the numerical calculation of the scalar power spectrum, the first step involves solving the background equations. With the background solutions $a(t)$ and $\phi(t)$ established, we construct the equations for the scalar perturbations, allowing us to numerically calculate the scalar power spectrum \cite{rojas:2022, tapia:2020}.

\section{The Starobinsky inflationary model}

The Starobinsky inflationary potential was introduced in the $80$'s by A. Starobinsky \cite{starobinsky:1980}, this potential is given by 

\begin{equation}
\label{V_Starobinsky}
V(\phi)=\dfrac{3}{4} M^2 \left(1-e^{-\sqrt{\sfrac{2}{3}}\,\phi}\right)^2,
\end{equation}
where $M=1.30 \times 10^{-5}$ \cite{mishra:2018} to fit the amplitude of the scalar power spectrum, and $\phi$ is the inflaton.

From Eqs. \eqref{epsilon}--\eqref{omega} with the potential given by Eq. \eqref{V_Starobinsky} we calculate analytically the slow--roll parameters for the Starobinsky inflationary model, and following the notation of Renzi \textit{et al.} \cite{renzi:2020} we found that they are given by:

\begin{eqnarray}
\label{epsilon_S}
\epsilon_\nu &=& \dfrac{4}{3}\dfrac{1}{\left(F_1-1\right)^2},\\
\label{eta_S}
\eta_\nu     &=& -\dfrac{4}{3}\dfrac{\left(F_1-2\right)}{\left(F_1-1\right)^2},\\
\xi_\nu      &=& \dfrac{16}{9}\dfrac{\left(F_1-4\right)}{\left(F_1-1\right)^3},\\
\omega_\nu   &=& -\dfrac{64}{27}\dfrac{\left(F_1-8\right)}{\left(F_1-1\right)^4},
\end{eqnarray}
where $F_1=e^{\sqrt{\frac{2}{3}}\phi}$.
Using Eq. \eqref{e-folds} we obtain the expression for $\phi_\text{ini}$,

\begin{equation}
\phi_\text{ini}=\dfrac{1}{6} \left[-3 \sqrt{6} e^{\sqrt{\frac{2}{3}} \phi_\text{end}}-4 \sqrt{6} N_e +6 \phi_\text{end}-3 \sqrt{6} W_{-1}\left(-e^{-e^{\sqrt{\frac{2}{3}} \phi_\text{end}}-\frac{4 Ne}{3}+\sqrt{\frac{2}{3}} \phi_\text{end}}\right)\right],
\end{equation}
where $W_{-1}$ is the Lambert function, and  $\phi_\text{end}$ is obtained from Eq. \eqref{epsilon_S} for making $\epsilon_\nu=1$,

\begin{equation}
\label{phiend_Starobinsky}
\phi_\text{end}=\sqrt{\dfrac{3}{2}} \ln \left[\dfrac{1}{3} \left(3+2\sqrt{3}\right)\right].
\end{equation}

In Fig. \ref{phi_Starobinsky}, we show the evolution of the scalar field $\phi(t)$ as a function of time $t$ for the Starobinsky inflationary model. The figure illustrates the oscillation of the scalar field $\phi(t)$ after the end of inflation.

\begin{figure}[th!]
\centering
\includegraphics[scale=0.50]{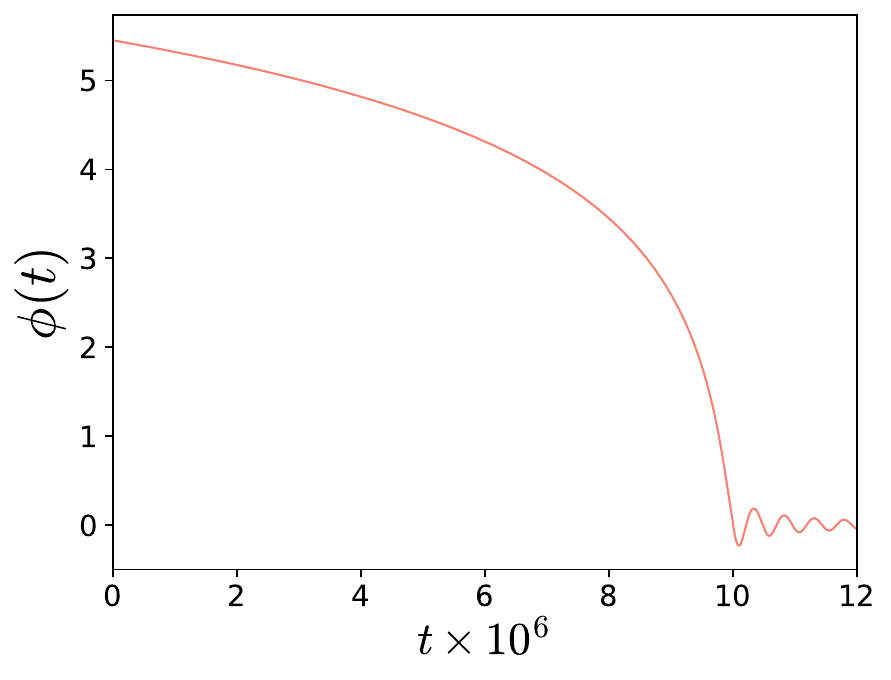}
\caption{Evolution of the scalar field $\phi(t)$ as a function of time $t$ of the Starobinsky inflationary model.}
\label{phi_Starobinsky}
\end{figure}

Using Eq. \eqref{r} we obtain that the value for the scalar--to--tensor ratio $r$ for the Starobinsky inflationary model is,

\begin{equation}
r=0.0029639.
\label{r_Starobinky}
\end{equation}

Figs. \ref{Starobinsky_epsilon} and \ref{Starobinsky_eta} shows that the slow--roll parameters \eqref{epsilon_S} and \eqref{eta_S} for the Starobinsky inflationary model are satisfied.

\begin{figure}[th!]
\begin{subfigure}[b]{0.49\textwidth}
\centering
\includegraphics[width=\textwidth]{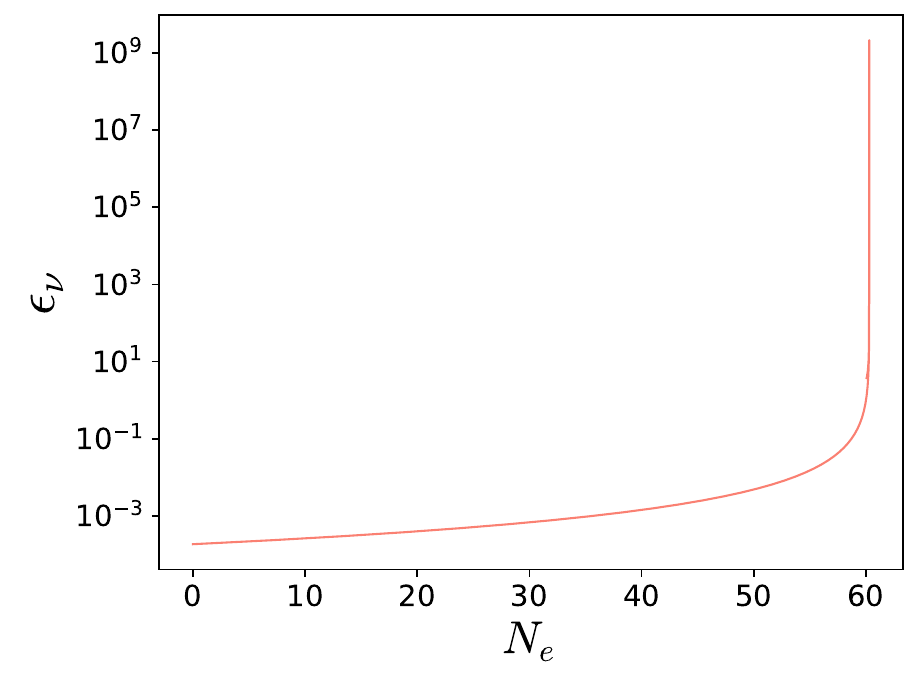}
\caption{}
\label{Starobinsky_epsilon}
\end{subfigure}
\begin{subfigure}[b]{0.49\textwidth}
\centering
\includegraphics[width=\textwidth]{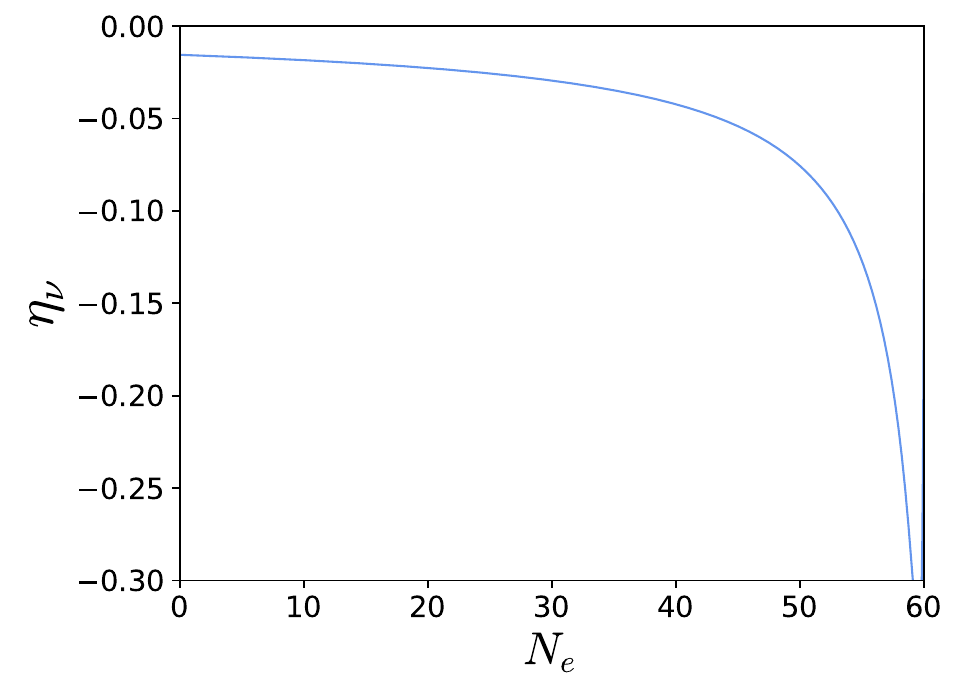}
\caption{}
\label{Starobinsky_eta}
\end{subfigure}
\caption{Behaviour of the slow--roll parameters $\epsilon_\nu$ and $\eta_\nu$ with respect to the number of e--folds $N_e$, described by the Starobinsky inflationary model.}
\label{slow-roll_Starobinsky}
\end{figure}

In Fig. \ref{Starobinsky_1} we present the CMB angular power spectrum for the Starobinsky inflationary model. 
The reproduced CMB angular power spectrum shows lower values of temperature fluctuations compared with Planck $2018$ results, as is observed in Fig. \ref{Starobinsky_3}.

In Fig. \ref{Starobinsky_2} is observed the large angular scales ($\ell \lesssim 90$) and its relative error is shown in Fig. \ref{error_Starobinsky_2}. At these scales the contribution to CMB
anisotropies is from perturbations that were superhorizon at recombination \cite{gorbunov2011introduction}.
In Fig. \ref{Starobinsky_3} is observed the intermediate angular scales ($90 < \ell \lesssim 900$) and its relative error is shown in Fig. \ref{error_Starobinsky_3}.
In Fig. \ref{Starobinsky_4} is observed the large angular scales ($\ell > 900$) and its relative error is shown in Fig. \ref{error_Starobinsky_4}.

\begin{figure}[th!]
\centering
\begin{subfigure}[b]{0.49\textwidth}
\centering
\includegraphics[width=\textwidth]{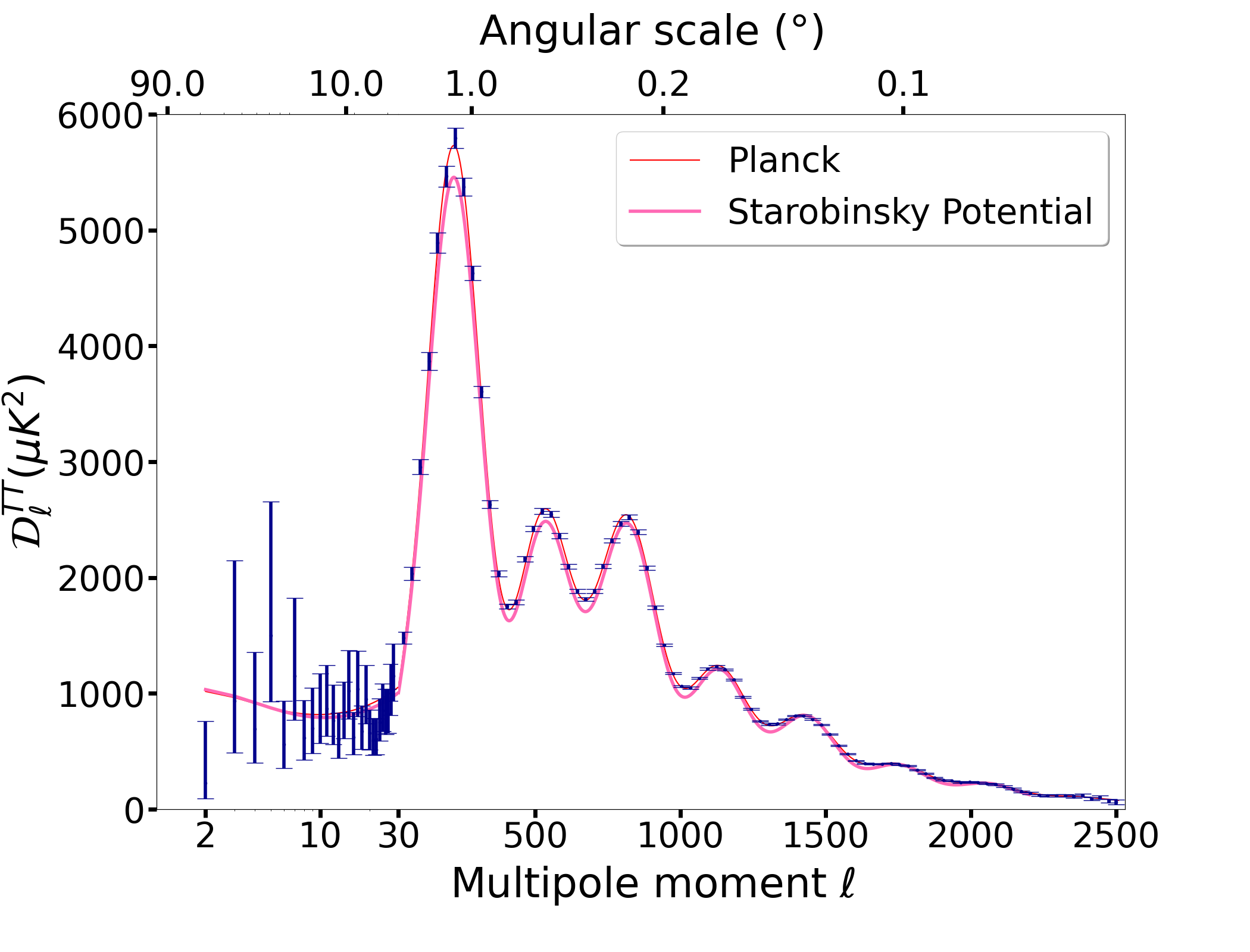}
\caption{}
\label{Starobinsky_1}
\end{subfigure}
\begin{subfigure}[b]{0.49\textwidth}
\centering
\includegraphics[width=\textwidth]{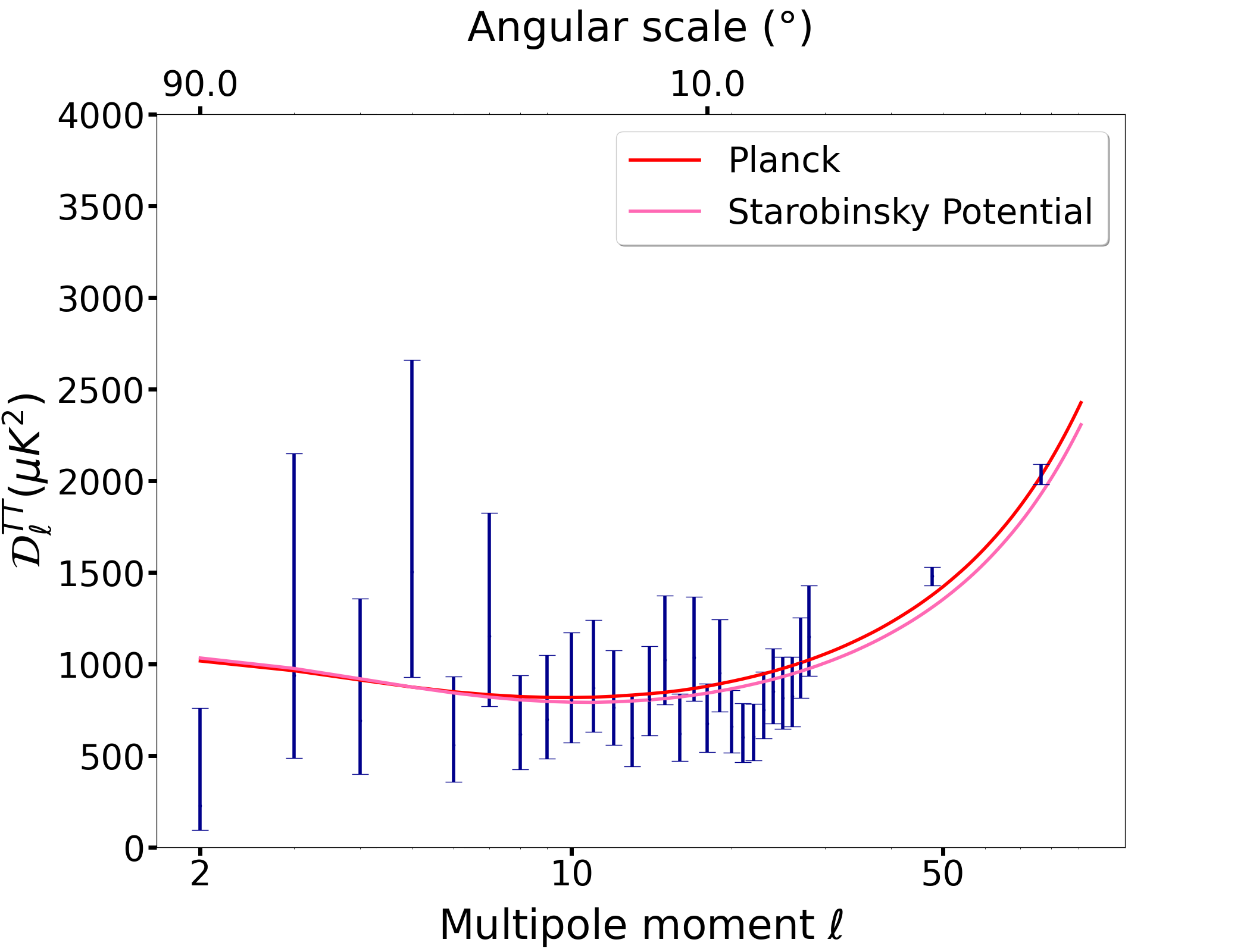}
\caption{}
\label{Starobinsky_2}
\centering
\end{subfigure}
\begin{subfigure}[c]{0.49\textwidth}
\includegraphics[width=\textwidth]{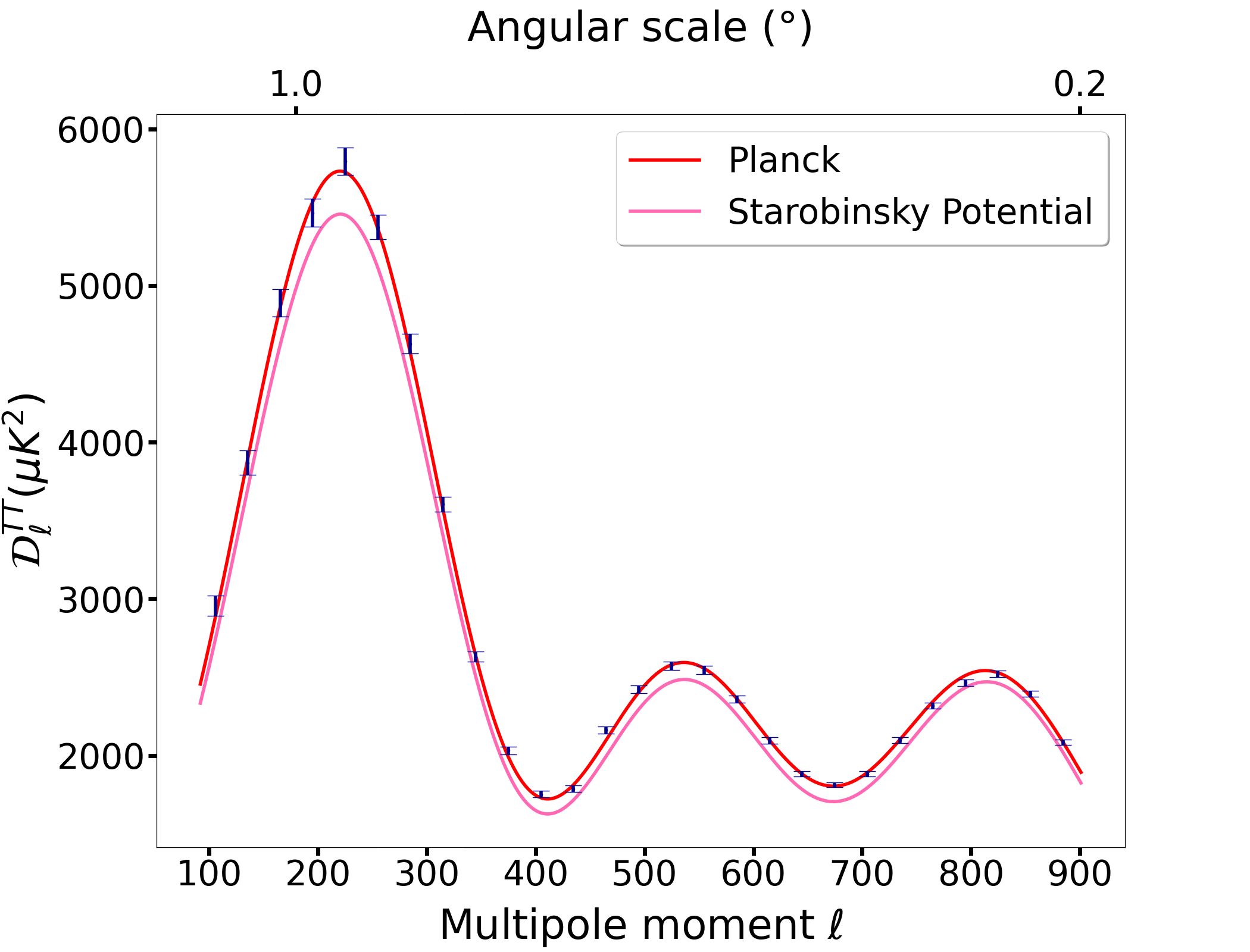}
\caption{}
\label{Starobinsky_3}
\end{subfigure}
\begin{subfigure}[d]{0.49\textwidth}
\centering
\includegraphics[width=\textwidth]{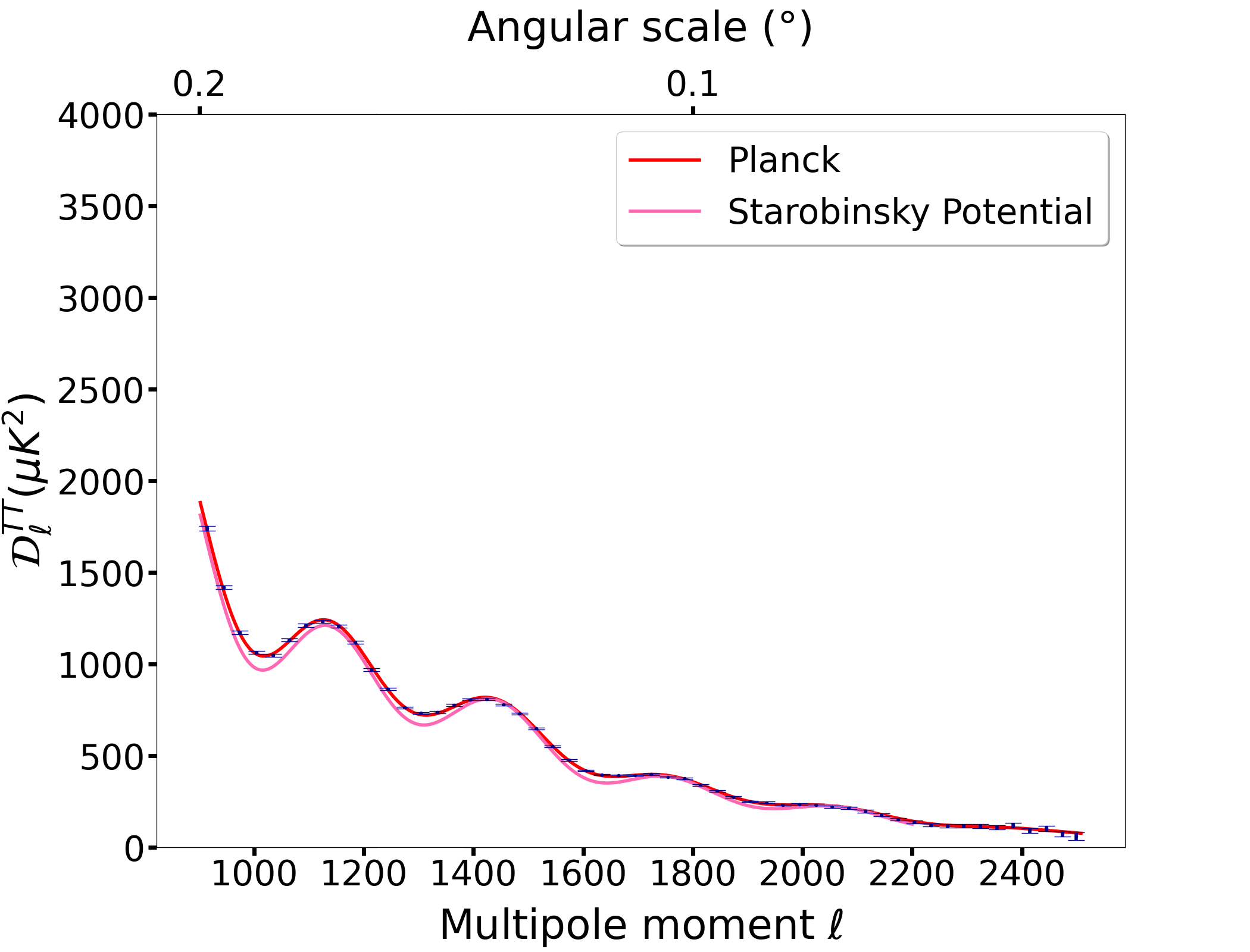}
\caption{}
\label{Starobinsky_4}
\end{subfigure}
\caption{(a) Angular power spectrum for the Starobinsky
inflationary model reproduced by \texttt{CAMB}, (b) large angular scales, (c) intermediate angular scales, and (d) small angular scales. Red  line: Planck $2018$ data with error bars in blue, pink line: \texttt{CAMB} result.}
\end{figure}

\begin{figure}[th!]
\centering
\begin{subfigure}[b]{0.50\textwidth}
\centering
\includegraphics[width=\textwidth]{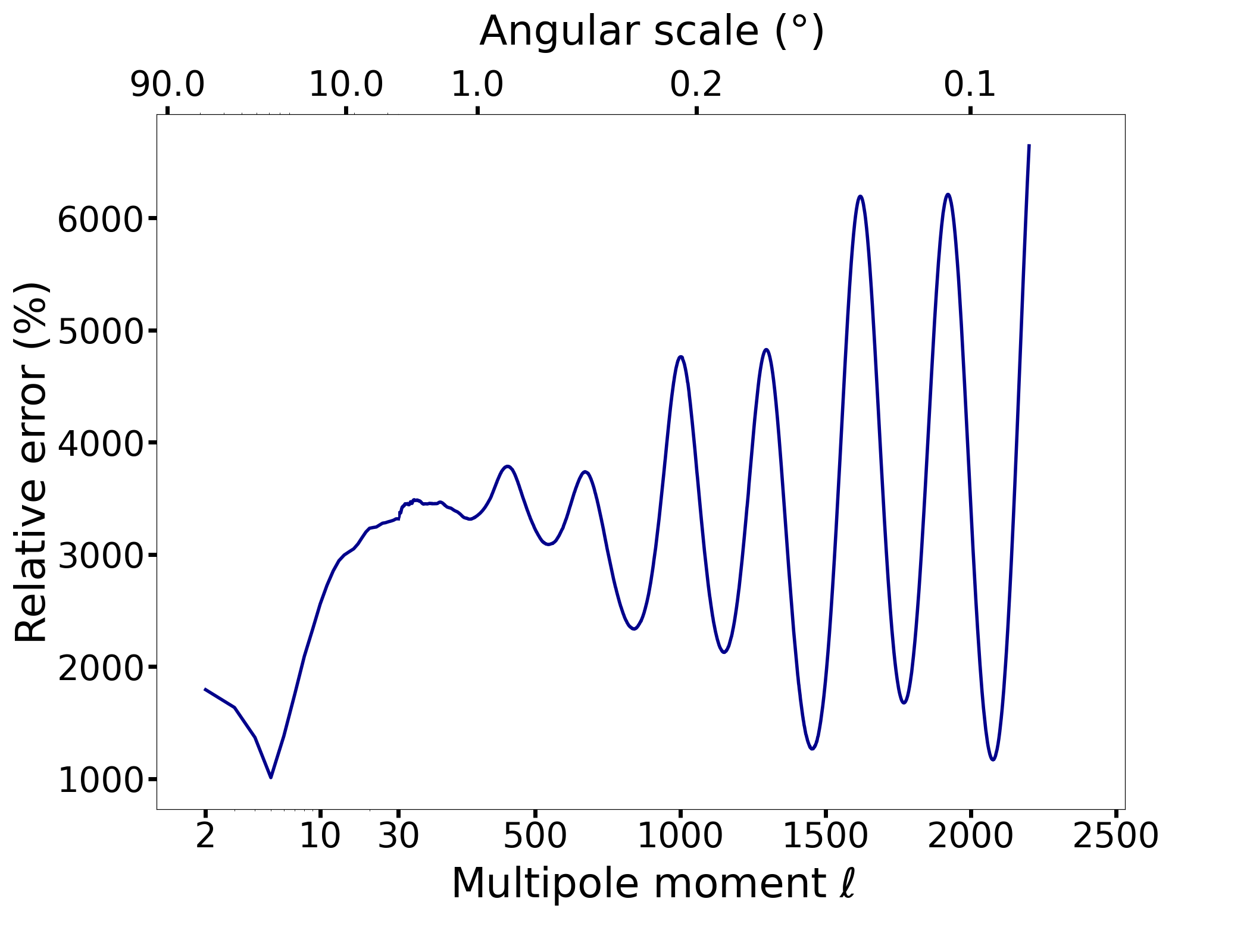}
\caption{}
\label{error_Starobinsky_1}
\end{subfigure}
\begin{subfigure}[b]{0.49\textwidth}
\centering
\includegraphics[width=\textwidth]{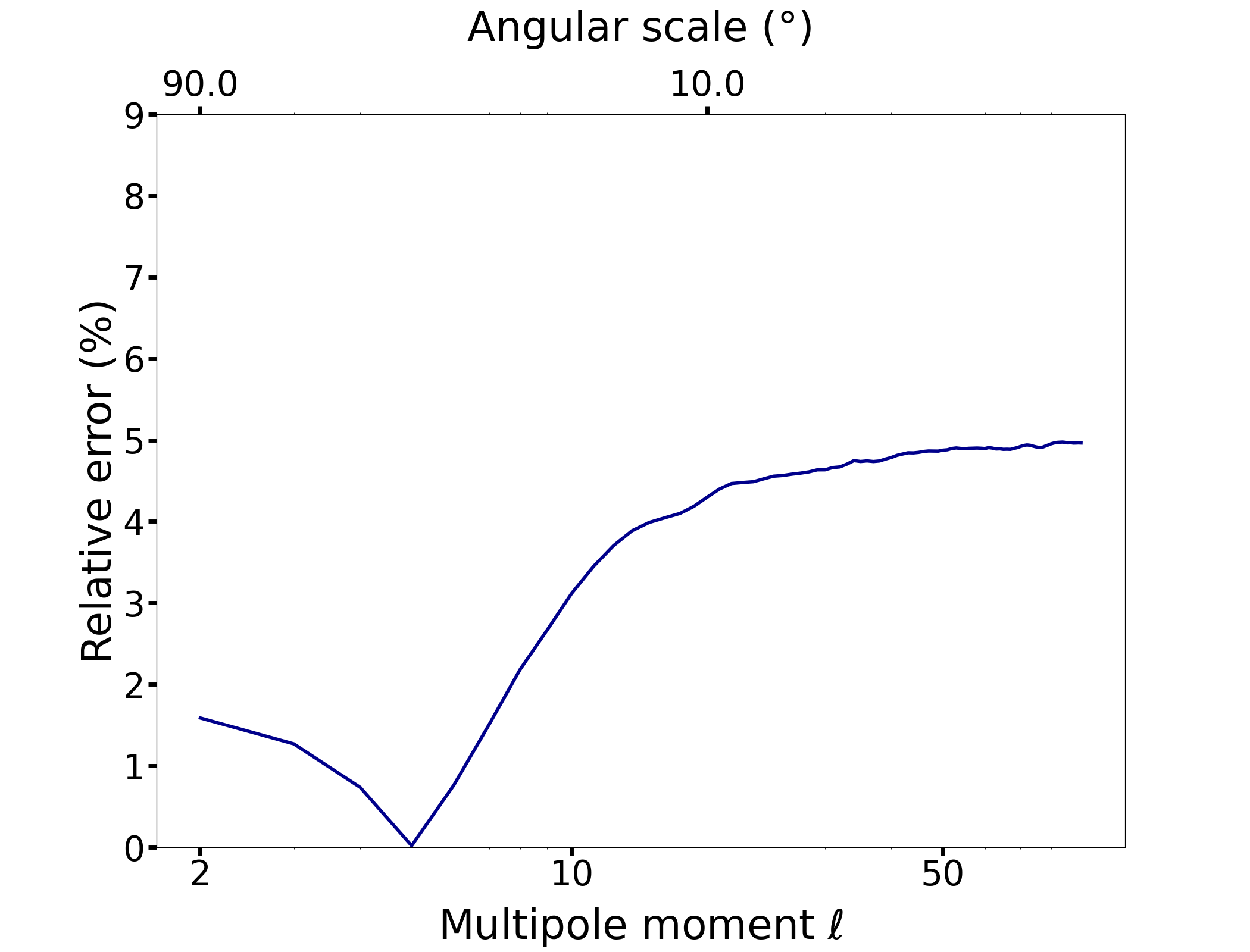}
\caption{}
\label{error_Starobinsky_2}
\centering
\end{subfigure}
\begin{subfigure}[c]{0.49\textwidth}
\includegraphics[width=\textwidth]{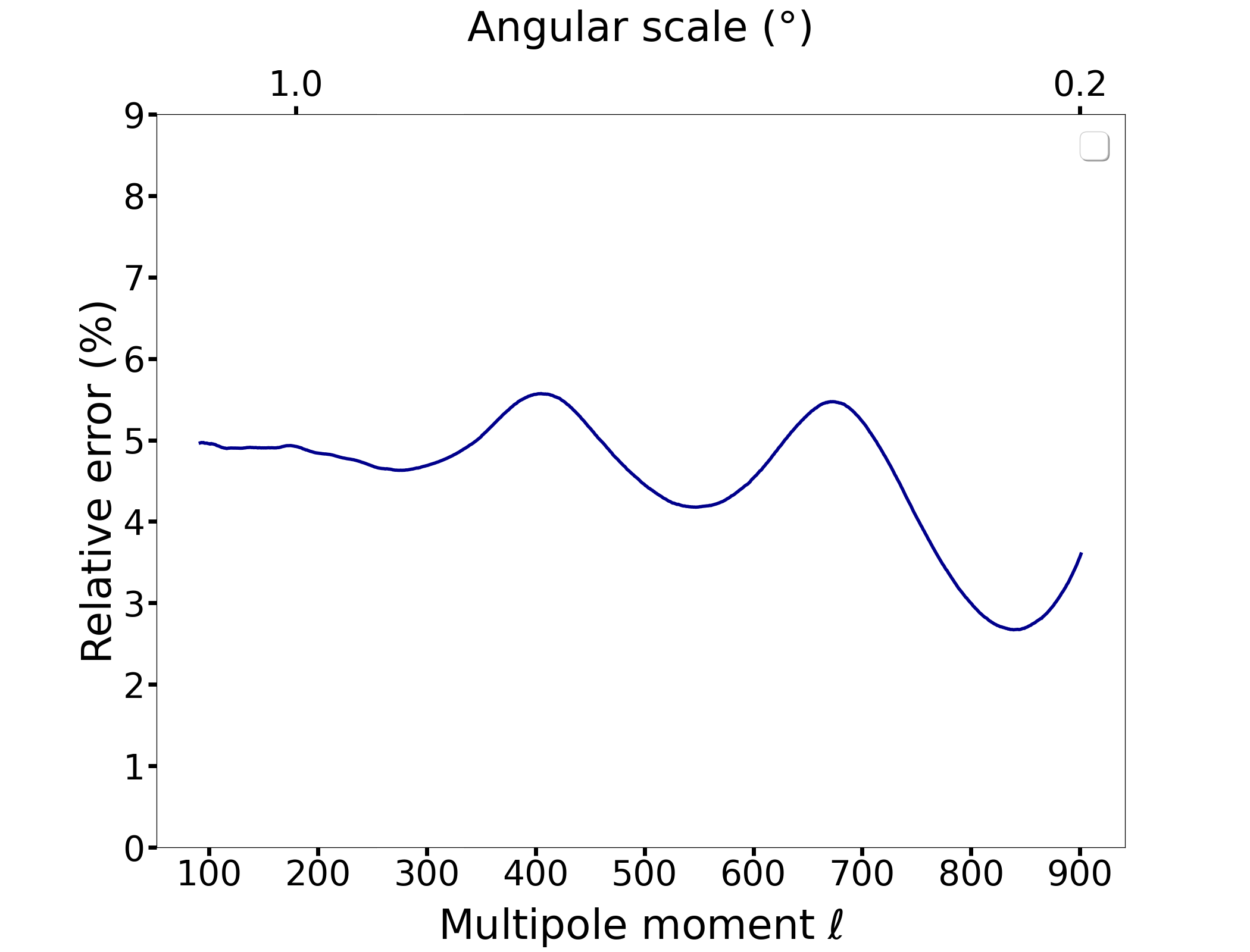}
\caption{}
\label{error_Starobinsky_3}
\end{subfigure}
\begin{subfigure}[d]{0.49\textwidth}
\centering
\includegraphics[width=\textwidth]{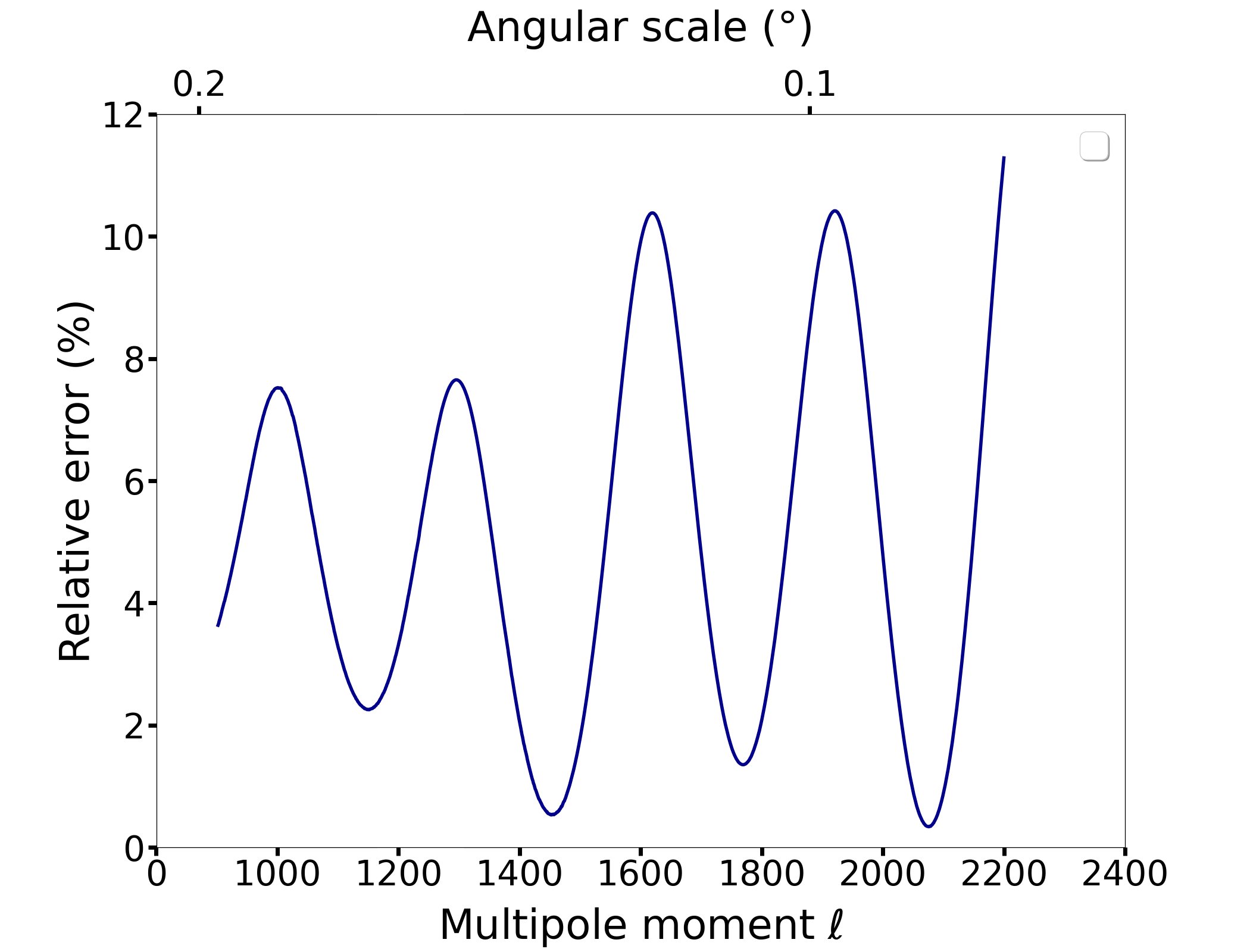}
\caption{}
\label{error_Starobinsky_4}
\end{subfigure}
\caption{(a) Error of the angular power spectrum for the Starobinsky inflationary model reproduced by \texttt{CAMB}, (b) large angular scales, (c) intermediate angular scales, and (d) small angular scales.}
\end{figure}

We can easily observe in the relative error plots that the \texttt{CAMB}--reproduced model aligns well with  Planck $2018$ observations, as indicated by the small errors. However, at smaller angular scales, results exhibit more variability, as evident from the oscillatory behavior of the errors in this section. To assess the overall balance of these errors, let us examine the L$2$ norm.

When considering all scales, this inflationary model has a L$2$ norm error of $94$ $\mu K^2$. At large, intermediate, and small angular scales, the L$2$ norm errors are $73$ $\mu K^2$, $146$ $\mu K^2$, and $37$ $\mu K^2$, respectively. Therefore, this model reproduces the observational CMB angular spectrum more accurately at smaller angular scales, but the error balance is less favorable at intermediate scales.

\subsection{The Acoustic Peaks}

In Table \ref{table1:Starobinsky}, the acoustic peaks found in the intermediate scale of the spectrum are listed, along with the troughs in this region. We observed that the positions on the multipole axis of the peaks are close to each other for both observed and computed values. Considering the errors of the observed values, all computed peaks, except the third one, agree with the Planck satellite. The percentage errors of the calculated multipole for peak $1$, trough $1$, peak $2$, trough $2$, and peak $3$ are $0.18$, $1.27$, $0.20$, $0.22$, and $0.52$, respectively.
However, the differences between the observed and computed amplitudes of the peaks and troughs appear to be greater. The percentage errors of the calculated amplitudes for peak $1$, trough $1$, peak $2$, trough $2$, and peak $3$ are $4.84$, $10.86$, $3.90$, $5.18$, and $1.89$, respectively. These errors indicate that, indeed, there is less agreement with the observations in the amplitude of the acoustic peaks than with their multipole.

\begin{table}[ht!]
\begin{center}
\makebox[\linewidth]{
\begin{tabular}{lcccc}
\toprule
& \multicolumn{2}{c}{\textbf{Starobinsky inflationary model}} & \multicolumn{2}{c}{\textbf{Planck Satellite}}\\
\textbf{Extremum} & \textbf{Multipole [$\ell$]} & \textbf{Amplitude [$\mu K^2$]} & \textbf{Multipole [$\ell$]} & \textbf{Amplitude [$\mu K^2$]}\\ 
\midrule
Peak $1$   & 221 & 5455.73 $\pm$ 0.05 & 220.6 $\pm$ 0.6 & 5733 $\pm$ 39 \\
Trough 1 & 411 & 1526.90 $\pm$ 0.11 & 416.3 $\pm$ 1.1 & 1713 $\pm$ 20 \\
Peak $2$   & 537 & 2485.11 $\pm$ 0.04 & 538.1 $\pm$ 1.3 & 2586 $\pm$ 23 \\
Trough 2 & 674 & 1705.74 $\pm$ 0.05 & 675.5 $\pm$ 1.2 & 1799 $\pm$ 14 \\
Peak $3$   & 814 & 2470.51 $\pm$ 0.03 & 809.8 $\pm$ 1.0 & 2518 $\pm$ 17 \\ 
\bottomrule      
\end{tabular}
}
\end{center}
\caption{Peaks and troughs of the CMB  TT  power spectra in the Acoustic Peak region recreated by the Starobinsky inflationary model and reported by Planck satellite.}
\label{table1:Starobinsky}
\end{table}

\subsection{Cosmological parameters}

In Table \ref{table2:Starobinsky}, some cosmological parameters theoretically calculated from the Starobinsky inflationary model are compared with their respective observational values from the Planck satellite. Considering the uncertainties, all the parameters are in agreement between their theoretical and observational values.

\begin{table}[ht!]
\begin{center}
\begin{tabular}{lccc}
\toprule
\textbf{Cosmological Parameter} & \textbf{Symbol} & \textbf{Starobinsky} & \textbf{Planck} \\\midrule
Age of Universe [Gyr] & Age              & $13.798 \pm 0.000$    & $13.797 \pm 0.023$   \\
Matter density        & $\Omega_m$         & $0.3158 \pm 0.0016$   & $0.3153 \pm 0.073$   \\
Baryon density        & $\Omega_b h^2$     & $0.02238 \pm  0.0004$ & $0.02237 \pm 0.0001$ \\
Dark energy density   & $\Omega_{\Lambda}$ & $0.6841 \pm 0.0009$   & $0.6847 \pm 0.0073$  \\
Scalar spectral index          & $n_\sca$              & $0.9678 \pm 0.0030$   & $0.9649 \pm 0.0042$  \\
\bottomrule
\end{tabular}
\end{center}
\caption{Cosmological parameters obtained from the Starobinsky inflationary model compared with the cosmological parameters reported by Planck $2018$ results.  }
\label{table2:Starobinsky}
\end{table}

The percentage errors for the age of the universe, matter density, baryon density, dark energy density, and scalar spectral index are $0.007$, $0.16$, $0.04$, $0.09$, and $0.2$, respectively. The low percentage errors demonstrate the goodness, accuracy, and precision that the Starobinsky inflationary model exhibits when predicting the observed cosmological parameters.

\newpage
\section{The generalized Starobinsky inflationary model}

The generalized Starobinsky inflationary model is given by \cite{renzi:2020,martin:2014b}, and it is a slight modification of the Starobinsky inflationary model,

\begin{equation}
\label{V_gStarobinsky}
V(\phi)= V_0 e^{-2 \sqrt{\frac{2}{3}}\phi} \left(e^{\sqrt{\frac{2}{3}}\phi}-1 \right)^{\frac{2p}{2p-1}},
\end{equation}
where $p$ is a real number close to unity, $\phi$ is the inflaton, and

\begin{equation}
V_0=6 \left(\dfrac{2p -1}{4p} \right) M^2 \left(\dfrac{1}{2p} \right)^{\frac{1}{2p-1}}.
\end{equation}

At $p=1$, equation \eqref{V_gStarobinsky} reduces to the Starobinsky inflationary  potential \cite{mishra:2018,martin:2014b}. The value of $M$ is fixed  in $M =1.30 \times 10^{-5}$  in order to obtain the parametrization of the amplitude for the scalar power spectrum at  the pivot scale $k=0.05$ Mpc$^{-1}$ in  the Starobinsky inflationary model $(p=1)$ \cite{canko:2020}. 

From Eqs. \eqref{epsilon}--\eqref{omega} using the potential given by Eq. \eqref{V_gStarobinsky}, we calculate the slow--roll parameters using a symbolic manipulation program, and following the notation of Renzi \cite{renzi:2020} we obtain that:

\vspace{-0.4cm}
\begin{eqnarray}
\label{epsilon_gStarobinsky}
\epsilon_\nu &=& \dfrac{4}{3 \left(2p-1\right)^2\left(F_1-1\right)^2}\Bigg[(1-2p)+(p-1)F_1\Bigg]^2,\\
\label{eta_gStarobinsky}
\nonumber
\eta_\nu     &=& \dfrac{4}{3 \left(2p-1\right)^2\left(F_1-1\right)^2}\Bigg[(8p^2-8p+2)+(-10 p^2+13p-4)F_1+(2 p^2-4p+2)F_1^2\Bigg],\\
\\
\nonumber
\xi_\nu      &=&  \dfrac{16}{9 \left(2p-1\right)^4\left(F_1-1\right)^4}\Bigg[ \left(64p^4-128 p^3+96 p^2-32 p +4 \right)\Bigg.\\
\nonumber
&+& \left(-168 p^4+380p^3-318 p^2+117p-16 \right) F_1 +\left(148 p^4-388p^3+373 p^2-156p+24\right) F_1^2\\
\nonumber
&+& \left(-48 p^2+150p^3-173p^2+87p-16  \right) F_1^3 +\left(4 p^4-16p^3+24p^2-16p+4\right) F_1^4,\\\\
\nonumber
\omega_\nu      &=&  \dfrac{64}{27 \left(2p-1\right)^6\left(F_1-1\right)^6}\Bigg[ \left( 512 p^6-1536 p^5+1920 p^4-1280 p^3+480 p^2-96 p +8 \right)\Bigg.\\
\nonumber
&+& \left(-2080 p^6+6736 p^5-9040 p^4+6440 p^3-2570 p^2+545 p -48 \right) F_1 \\
\nonumber
&+& \left(3328 p^6-11744 p^5+17088 p^4-13136 p^3+5632 p^2-1278 p +120  \right) F_1^2\\
\nonumber
&+& \left(-2632p^6+10236 p^5-16350 p^4+13741 p^3-6414p^2+1578p-160\right) F_1^3\\
\nonumber
&+& \left(1048 p^6-4536 p^5+8070 p^4-7552 p^3+3920p^2-1070p+120\right) F_1^4\\
\nonumber
&+& \left(-184 p^6+900p^5-1822 p^4+1953 p^3-1168p^2+369p-48\right) F_1^5\\
&+& \Bigg. \left(-2632p^6+10236 p^5-16350p^4+13741p^3-6414p^2+1578p-160\right) F_1^6\Bigg].
\end{eqnarray}

For this model, we need to calculate the values of $\phi_\text{ini}$ numerically using Eq. \eqref{e-folds}, and $\phi_\text{end}$ from Eq. \eqref{eta_gStarobinsky} when $\epsilon_\nu =1$, which becomes:

\begin{equation}
\label{phiend_gStarobinsky}
\phi_\text{end}=\sqrt{\dfrac{3}{2}}\ln\left[\dfrac{1-4p^2-2\sqrt{3\left(p^2-4p^3+4p^4\right)}}{1+4p^2-8p^2}\right],
\end{equation}
where Eq. \eqref{phiend_gStarobinsky} reduces to Eq. \eqref{phiend_Starobinsky} for the Starobinsky inflationary model when $p=1$.  In Fig. \ref{phi_gStarobinsky}, we show the evolution of the scalar field $\phi(t)$ as a function of time $t$ of the  generalized Starobinsky inflationary model considering $p=1.0004$. The figure illustrates the oscillation of the scalar field $\phi(t)$ after the end of inflation.

\begin{figure}[ht!]
\centering
\includegraphics[scale=0.5]{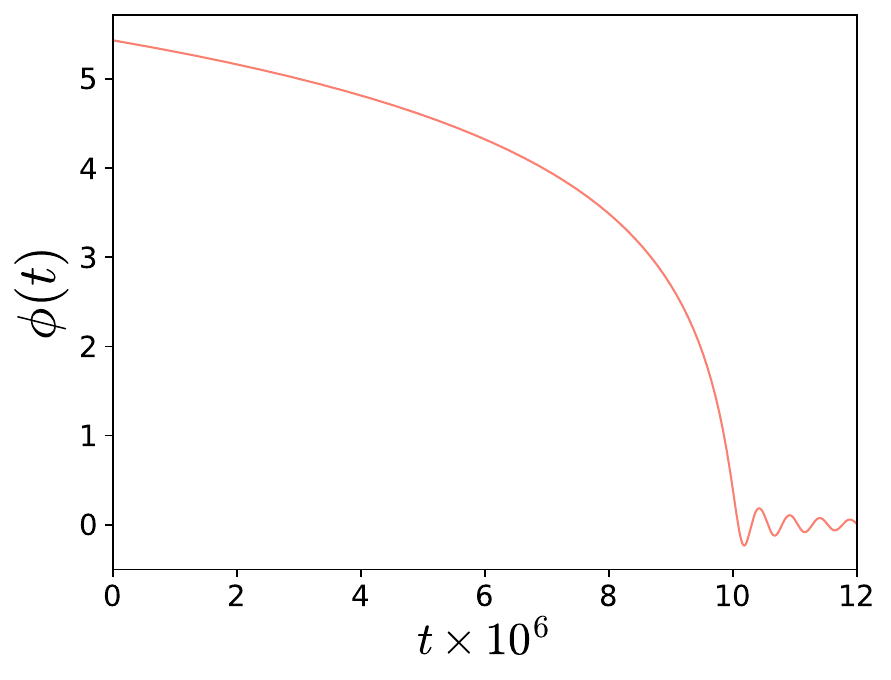}
\caption{Evolution of the scalar field $\phi(t)$ as a function of time $t$ of the generalized Starobinsky inflationary model for $p=1.0004$.}
\label{phi_gStarobinsky}
\end{figure}

Using Eq. \eqref{r} we obtain that the value for the scalar--to--tensor ratio $r$ for the generalized Starobinsky inflationary model for each considered case which is shown in Table \ref{r:gStarobinsky}. From table \ref{r:gStarobinsky} we can observed that for $p=1$ we recover the result of the Starobinsky inflationary model.

\begin{table}[th!]
\begin{center}
\begin{tabular}{cc}
\toprule
$p$ & $r$\\
\midrule
0.995 & 0.00449759\\
0.996 & 0.00404992\\
0.997 & 0.00381766\\
0.998 & 0.00351217\\
1.000 & 0.0029639\\
1.0004 & 0.00286365\\
1.001 & 0.00271881\\
1.002 & 0.00249167\\
1.003 & 0.00228136\\
1.004 & 0.00208686\\
1.005 & 0.00190719\\
\bottomrule
\end{tabular}
\end{center}
\caption{Value of the scalar--to--tensor ratio $r$ for each value of $p$ considered in the generalized Starobinsky inflationary model.}
\label{r:gStarobinsky}
\end{table}

Fig. \ref{gStarobinsky_epsilon}  and \ref{gStarobinsky_eta} show that the slow--roll parameters \eqref{epsilon_gStarobinsky} and \eqref{eta_gStarobinsky} for the generalized Starobinsky inflationary model are satisfied.

\begin{figure}[th!]
\begin{subfigure}[b]{0.5\textwidth}
\centering
\includegraphics[width=\textwidth]{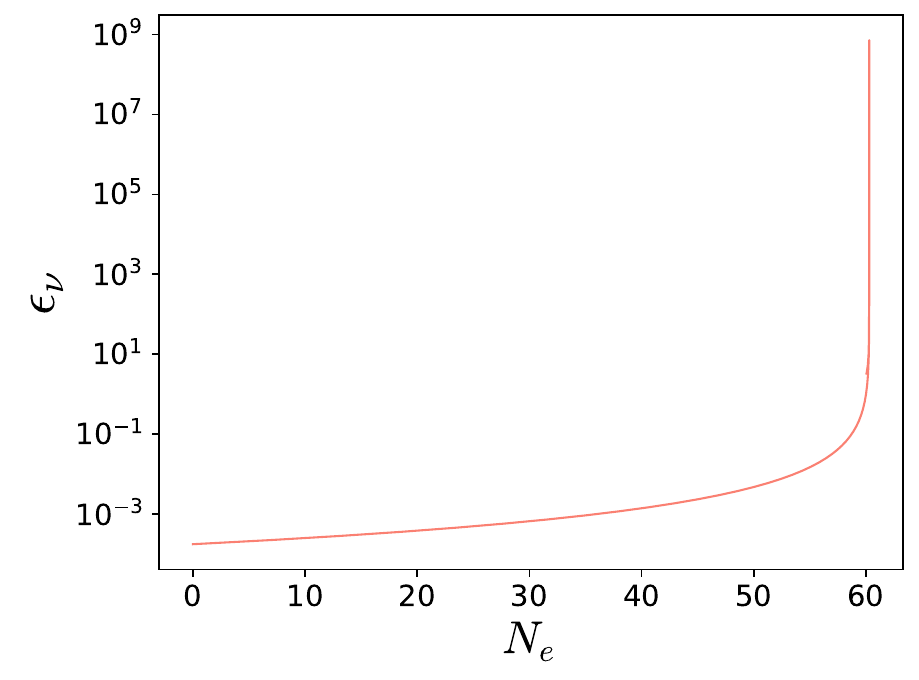}
\caption{}
\label{gStarobinsky_epsilon}
\end{subfigure}
\begin{subfigure}[b]{0.5\textwidth}
\centering
\includegraphics[width=\textwidth]{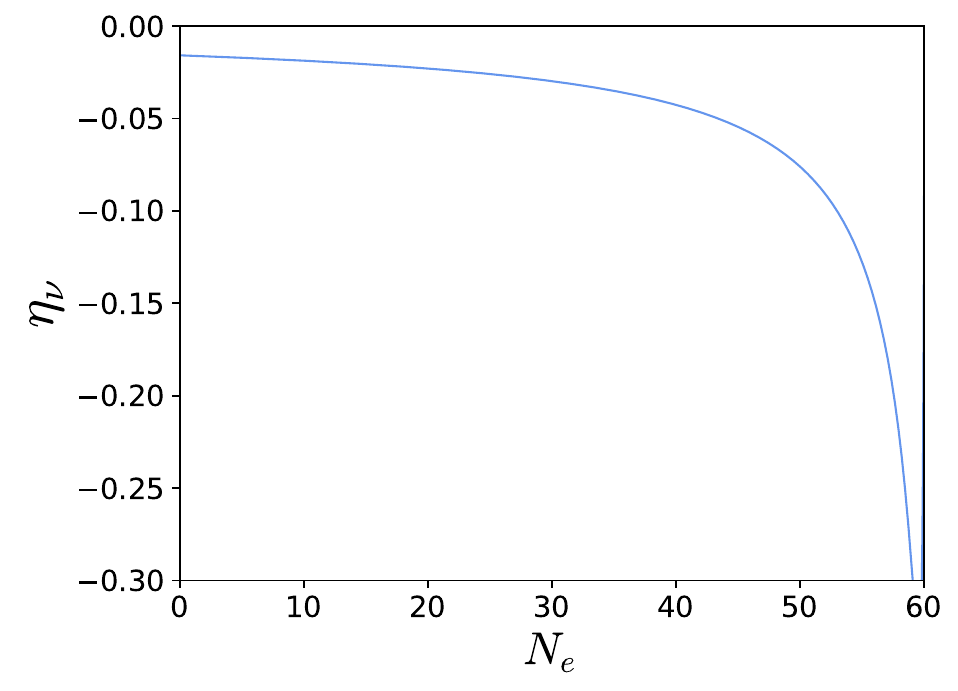}
\caption{}
\label{gStarobinsky_eta}
\end{subfigure}
\caption{Behaviour of the slow--roll parameters $\epsilon_\nu$ and $\eta_\nu$ with respect to the number of e--folds $N_e$, described by the generalized Starobinsky inflationary model for $p=1.0004$.}
\end{figure}

Fig. \ref{gStarobinsky_1} shows the CMB angular spectrum reproduced in \texttt{CAMB} for multiple values of the parameter $p$ in comparison to Planck $2018$ data. It is evident that for $p>1$, this parameter shifts the spectrum upwards, whereas for $p<1$ it shifts it downwards in comparison to the Starobinsky inflationary model ($p=1$). As the CMB reproduced by the Starobinsky inflationary model was under the observational data, we expect the optimal value of $p$ to be a little bit greater than $1$ to approach the observational curve.

Performing the \texttt{CAMB} reproduction of the CMB for various values of $p$, we found that the value that gives us the best tensor--to--scalar ratio $r$ and the scalar spectral index $n_\sca$, compared to the ones reported by Planck $2018$ results, is $1.0004 \pm 0.0001$, which is greater than $1$ as predicted before.  Now, let us plot it.

\begin{figure}[th!]
\includegraphics[width=\textwidth]{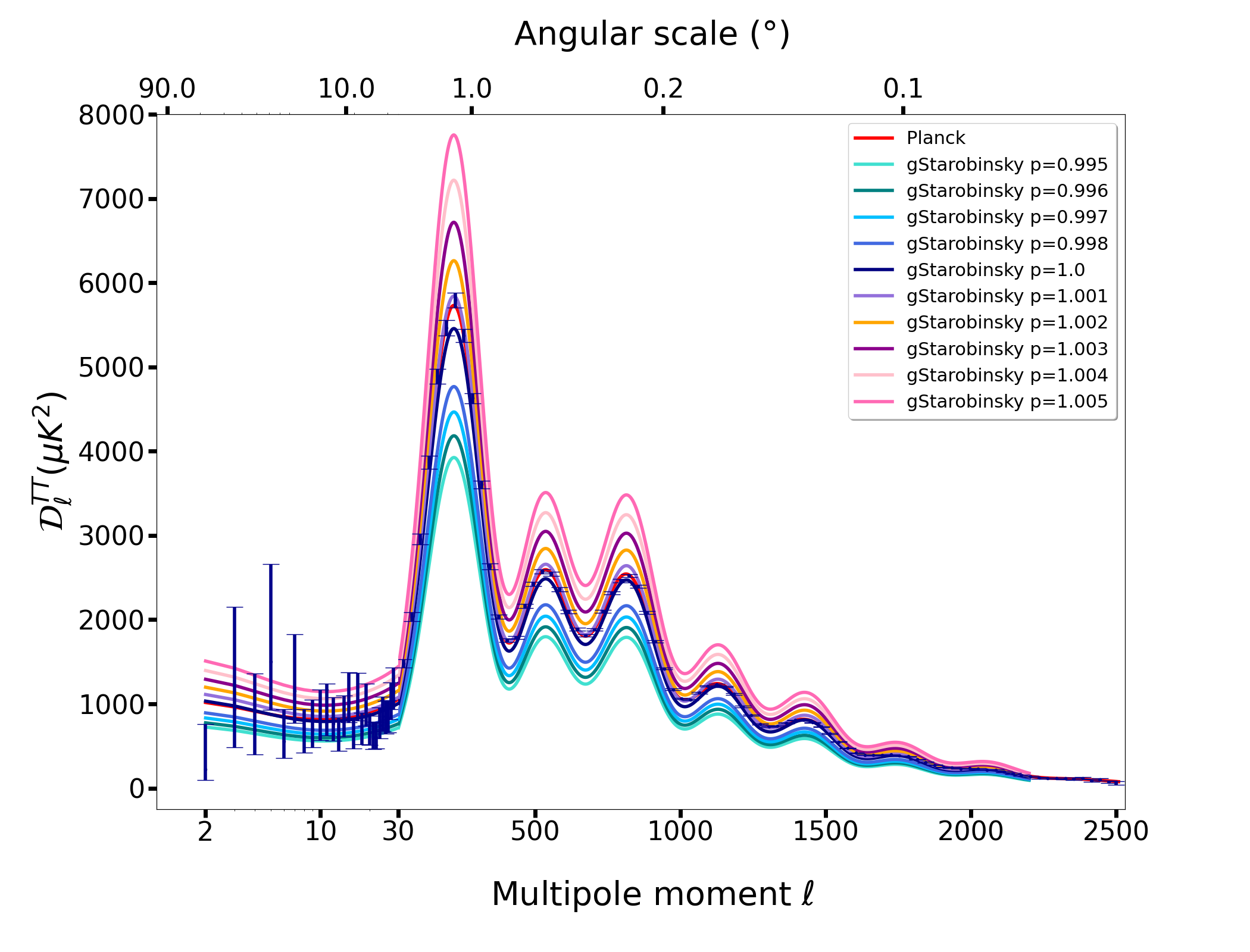}
\caption{Angular power spectrum for the generalized Starobinsky inflationary model reproduced by \texttt{CAMB} for different values of $p$. Red solid line: Planck $2018$ data.}
\label{gStarobinsky_1}
\end{figure}


In Fig. \ref{gStarobinsky_3} is observed the large angular scales ($\ell \lesssim 90$) at $p=1.0004$ and its relative error is shown in Fig. \ref{error_gStarobinsky_3}. 
In Fig. \ref{gStarobinsky_4} is observed the intermediate angular scales ($90 < \ell \lesssim 900$) at $p=1.0004$ and its relative error is shown in Fig. \ref{error_gStarobinsky_4}.
In Fig. \ref{gStarobinsky_5} is observed the large angular scales ($\ell > 900$) at $p=1.0004$ and its relative error is shown in Fig. \ref{error_gStarobinsky_5}. 

\begin{figure}[th!]
\centering
\begin{subfigure}[b]{0.50\textwidth}
\centering
\includegraphics[width=\textwidth]{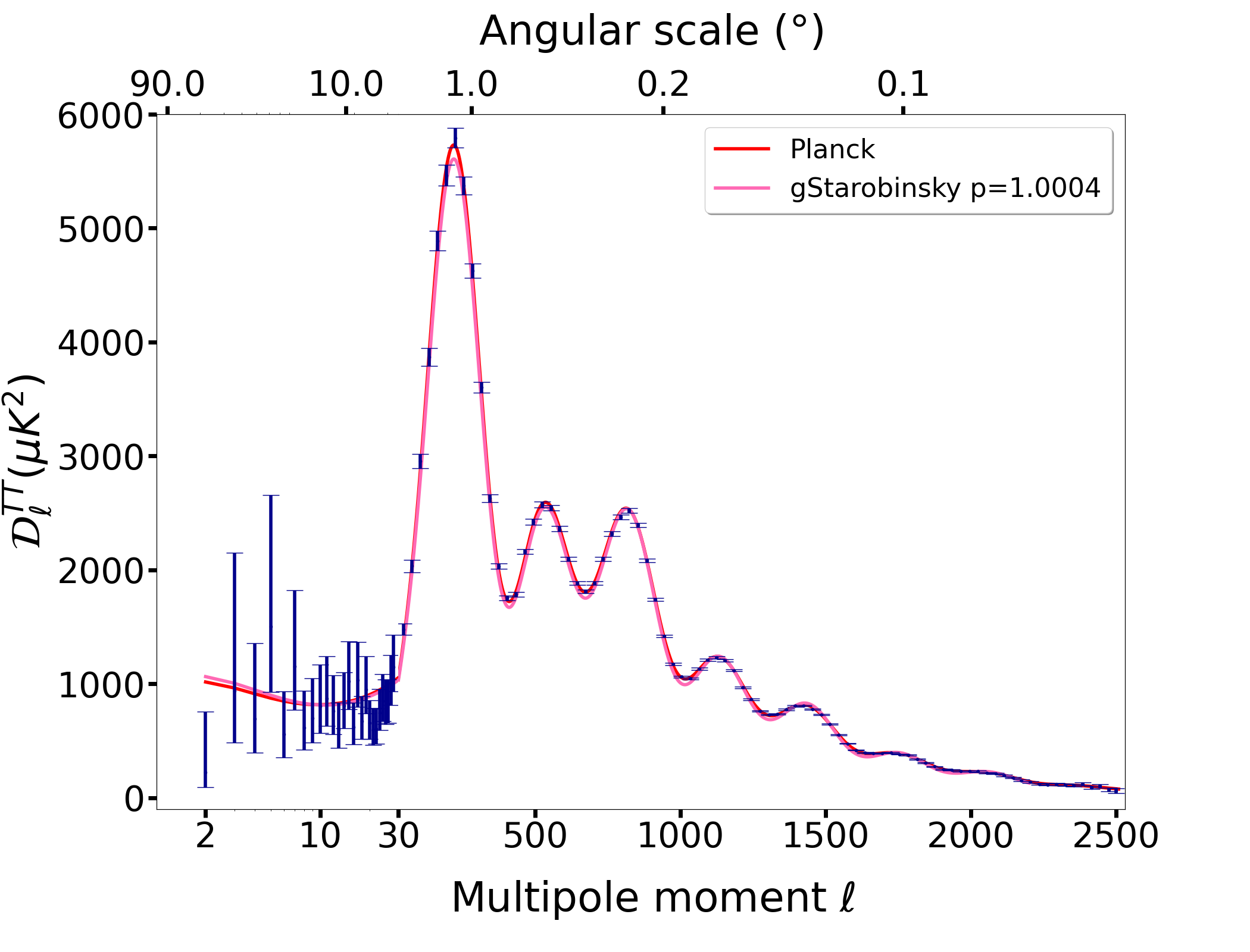}
\caption{}
\label{gStarobinsky_2}
\end{subfigure}
\begin{subfigure}[b]{0.49\textwidth}
\centering
\includegraphics[width=\textwidth]{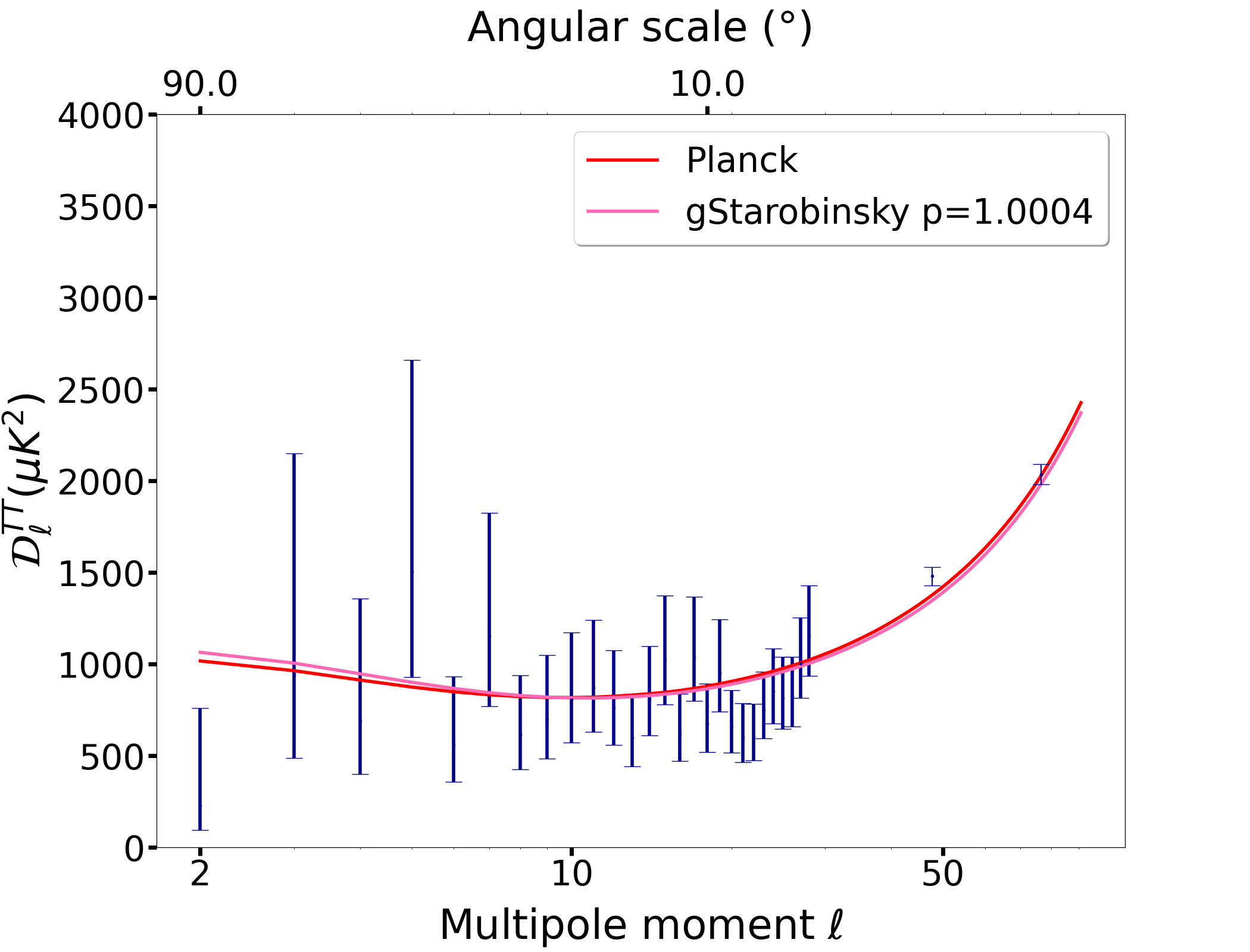}
\caption{}
\label{gStarobinsky_3}
\centering
\end{subfigure}
\begin{subfigure}[c]{0.49\textwidth}
\includegraphics[width=\textwidth]{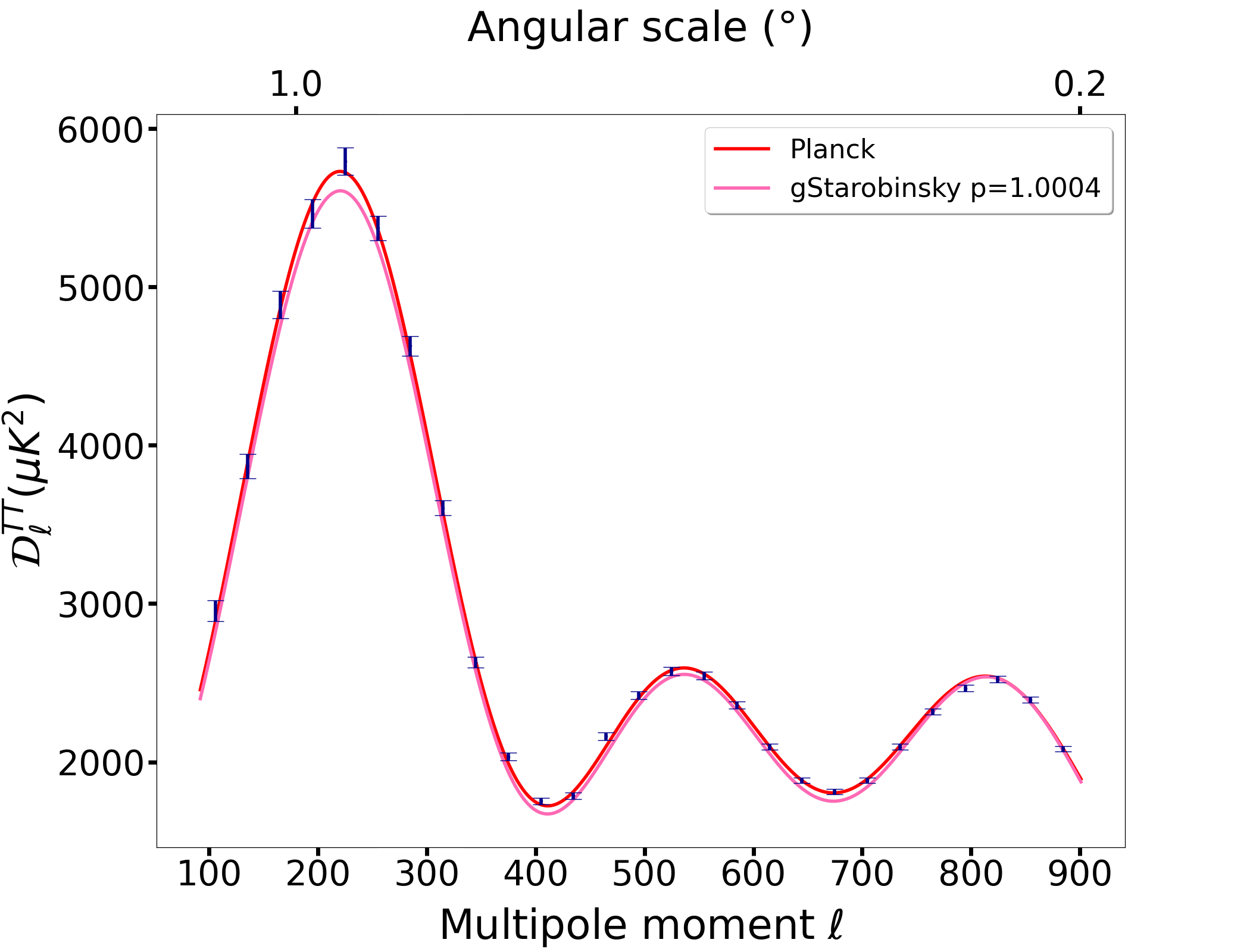}
\caption{}
\label{gStarobinsky_4}
\end{subfigure}
\begin{subfigure}[d]{0.49\textwidth}
\centering
\includegraphics[width=\textwidth]{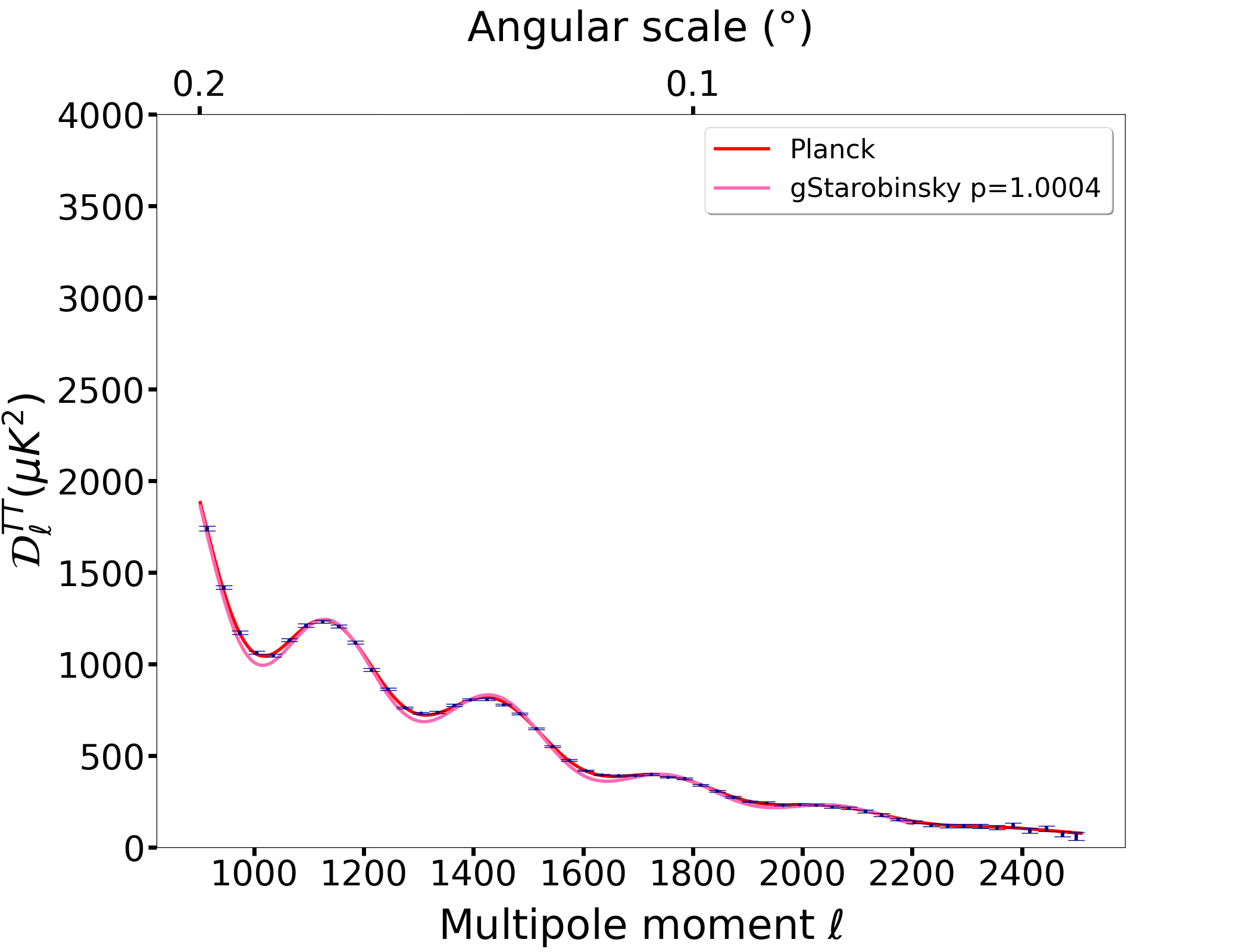}
\caption{}
\label{gStarobinsky_5}
\end{subfigure}
\caption{(a) Angular power spectrum for the generalized Starobinsky inflationary model  reproduced by \texttt{CAMB} for $p=1.0004$, (b) large angular scales, (c) intermediate angular scales, and (d) small angular scales. Red  line: Planck $2018$ data with error bars in blue, pink line: \texttt{CAMB} result.}
\end{figure}

\begin{figure}[th!]
\centering
\begin{subfigure}[b]{0.50\textwidth}
\centering
\includegraphics[width=\textwidth]{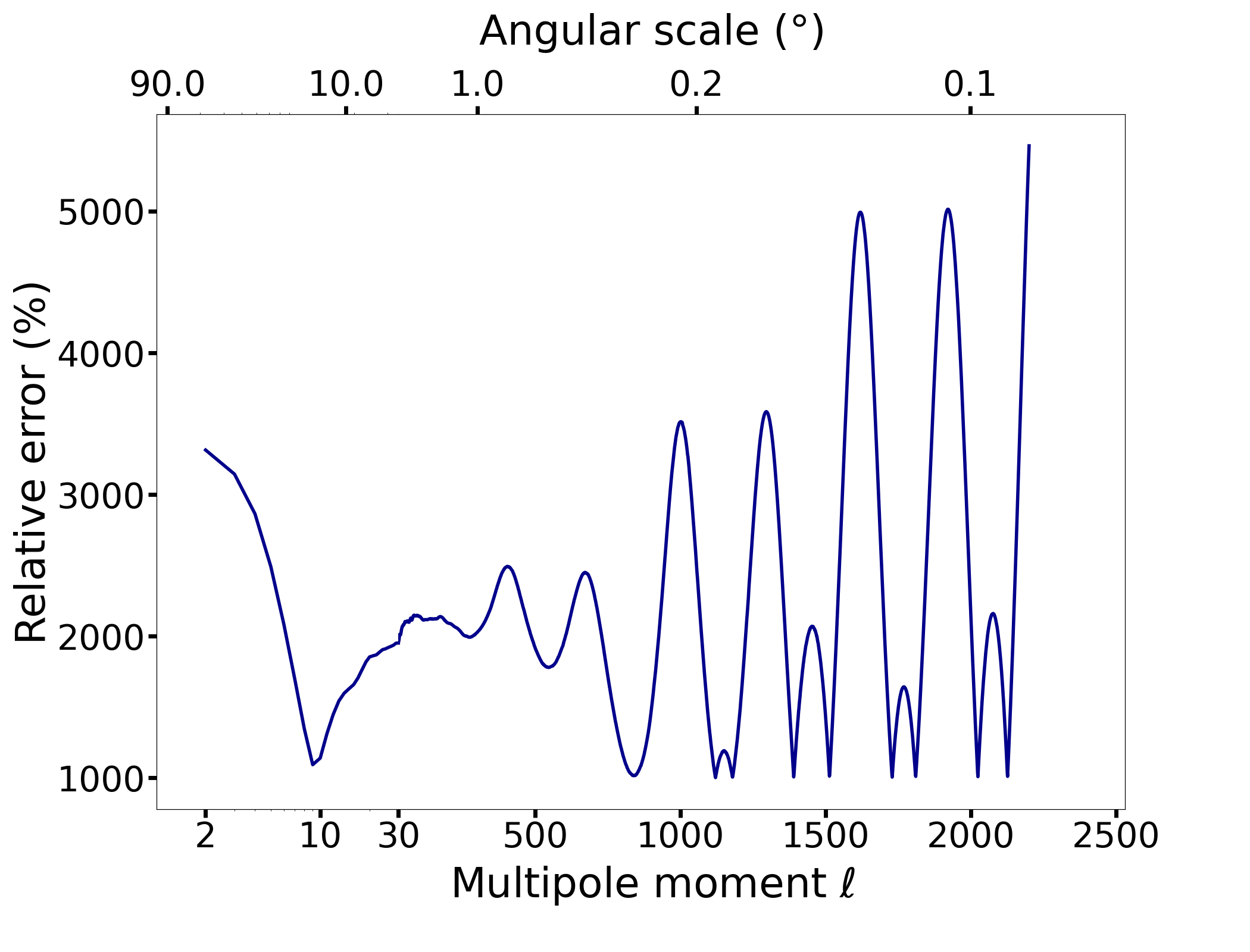}
\caption{}
\label{error_gStarobinsky_2}
\end{subfigure}
\begin{subfigure}[b]{0.49\textwidth}
\centering
\includegraphics[width=\textwidth]{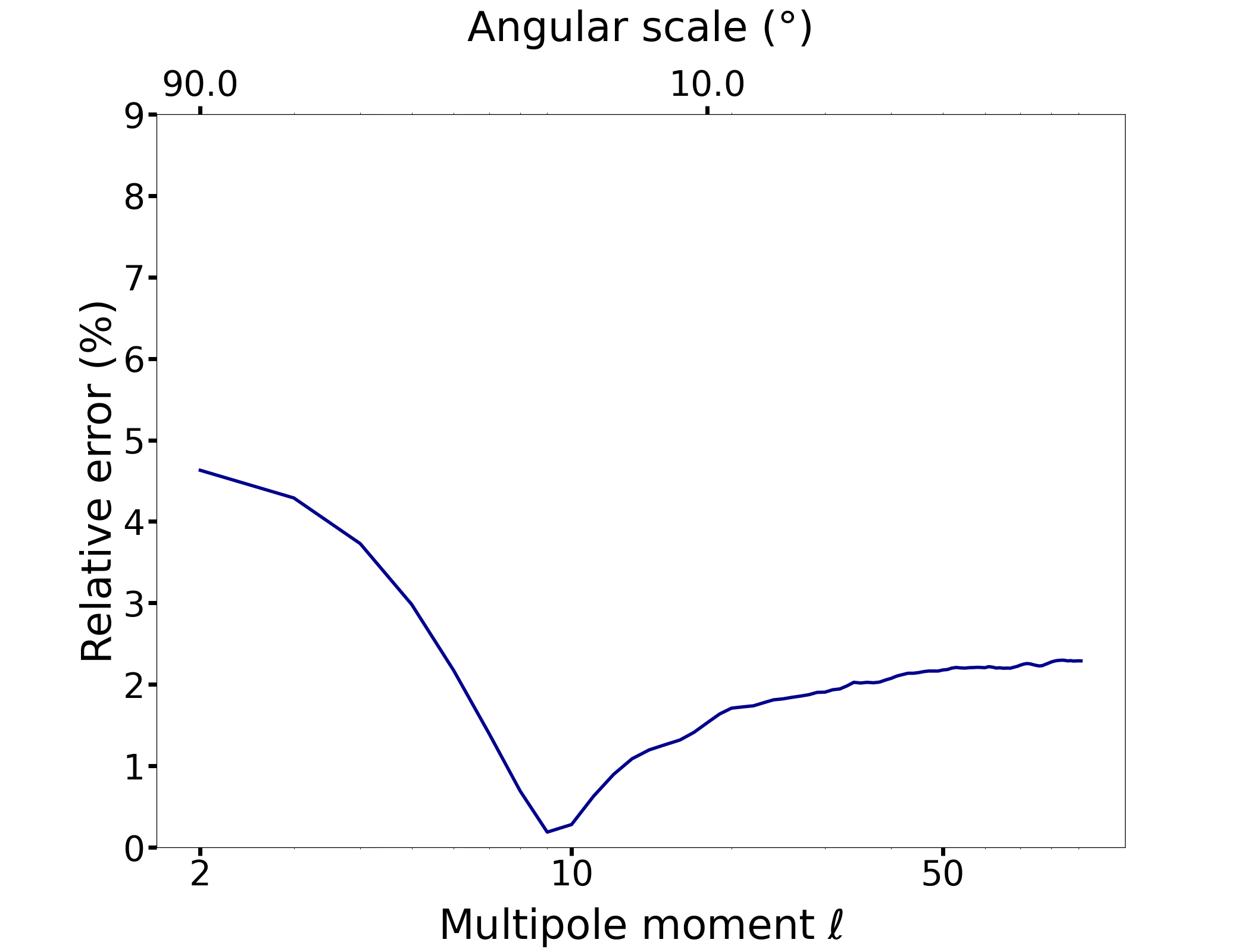}
\caption{}
\label{error_gStarobinsky_3}
\centering
\end{subfigure}
\begin{subfigure}[c]{0.49\textwidth}
\includegraphics[width=\textwidth]{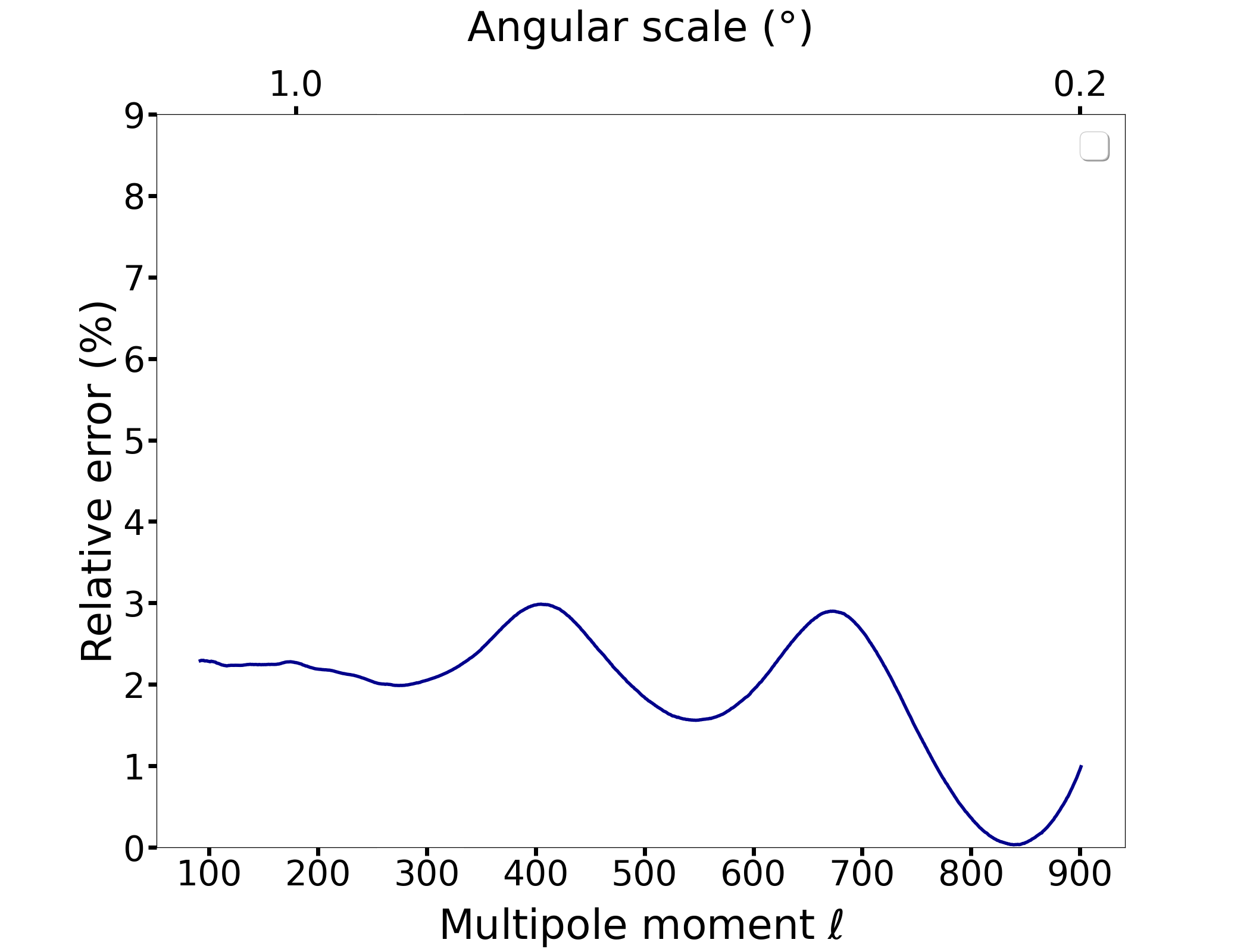}
\caption{}
\label{error_gStarobinsky_4}
\end{subfigure}
\begin{subfigure}[d]{0.49\textwidth}
\centering
\includegraphics[width=\textwidth]{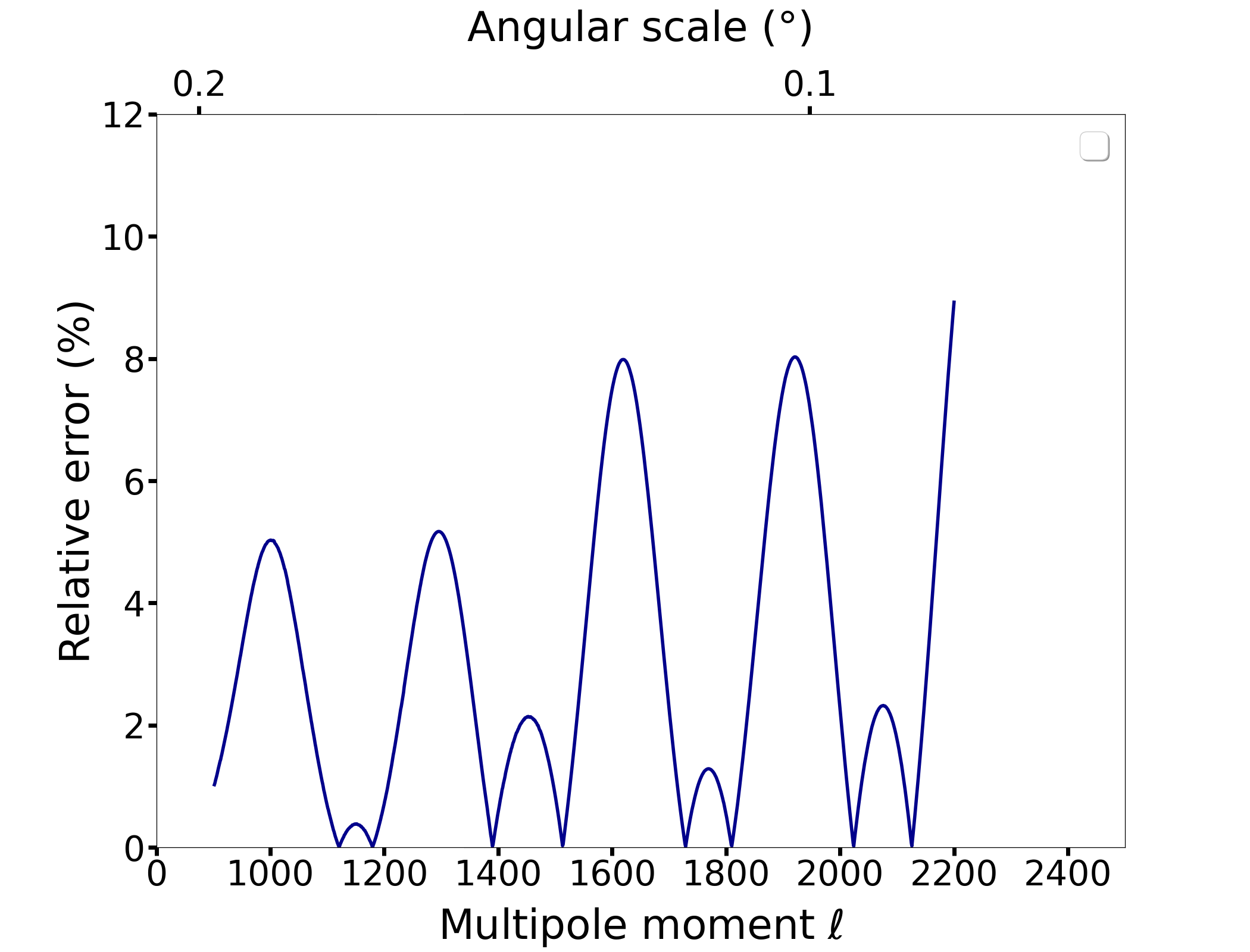}
\caption{}
\label{error_gStarobinsky_5}
\end{subfigure}
\caption{(a) Error of the angular power spectrum for the generalized Starobinsky
inflationary model  reproduced by \texttt{CAMB} for $p=1.0004$, (b) large angular scales, (c) intermediate angular scales, and (d) small angular scales.}
\end{figure}

For $p=1.0004$, this \texttt{CAMB}--reproduced model aligns well with Planck $2018$ observations, even surpassing the Starobinsky inflationary model, as evidenced by the smaller errors it exhibits in comparison. Variability in the results persists at small angular scales as in the previous model. To assess the overall balance of these errors, let us examine the L$2$ norm.

When considering all scales, this inflationary model has an L$2$ norm error of $43$ $\mu K^2$. At large, intermediate, and small angular scales, the L$2$ norm errors are $34$ $\mu K^2$, $65$ $\mu K^2$, and $22$ $\mu K^2$, respectively. Therefore, similar to the Starobinsky inflationary model, the generalized Starobinsky inflationary model reproduces the observational CMB angular spectrum more accurately at smaller angular scales, but the error balance is less favorable at intermediate scales. However, it significantly outperforms the Starobinsky inflationary model in terms of both the relative error size and the error balance.

\subsection{The Acoustic Peaks}

In Table \ref{table1:gStarobinsky}, we present the acoustic peaks identified in the intermediate range of the spectrum, together with the corresponding troughs. Our observations reveal close proximity in the positions along the multipole axis between the observed and calculated values. Considering the uncertainties associated with the observed data, all computed peaks, with the exception of the third one, align well with the Planck satellite. The percentage errors associated with the calculated multipole values for peak $1$, trough $1$, peak $2$, trough $2$, and peak $3$ are $0.18$, $1,27$, $0.20$, $0.22$, and $0.52$, respectively. This errors are exactly the same of those reported by the Starobinsky inflationary model.

\begin{table}[ht!]
\begin{center}
\makebox[\linewidth]{
\begin{tabular}{lcccc}
\toprule
& \multicolumn{2}{c}{\textbf{Generalized Starobinsky inflationary model}} & \multicolumn{2}{c}{\textbf{Planck Satellite}}\\
\textbf{Extremum} & \textbf{Multipole [$\ell$]} & \textbf{Amplitude [$\mu K^2$]} & \textbf{Multipole [$\ell$]} & \textbf{Amplitude [$\mu K^2$]}\\ 
\midrule
Peak $1$   & 221 & 5607.40 $\pm$ 0.02 & 220.6 $\pm$ 0.6 & 5733 $\pm$ 39 \\
Trough $1$ & 411 & 1671.43 $\pm$ 0.03 & 416.3 $\pm$ 1.1 & 1713 $\pm$ 20 \\
Peak $2$   & 537 & 2552.95 $\pm$ 0.01 & 538.1 $\pm$ 1.3 & 2586 $\pm$ 23 \\
Trough $2$ & 674 & 1752.16 $\pm$ 0.03 & 675.5 $\pm$ 1.2 & 1799 $\pm$ 14 \\
Peak $3$   & 814 & 2537.55 $\pm$ 0.00 & 809.8 $\pm$ 1.0 & 2518 $\pm$ 17 \\ 
\bottomrule     
\end{tabular}
}
\end{center}
\caption{Peaks and troughs of the CMB  TT  power spectra in the Acoustic Peak region recreated by the generalized Starobinsky inflationary model for $p=1.0004$ and reported by Planck satellite.}
\label{table1:gStarobinsky}
\end{table}

On the other hand, the percentage errors for the calculated amplitudes of peak $1$, trough $1$, peak $2$, trough $2$, and peak $3$ are $2.20$, $2.43$, $1.28$, $2.60$, and $0.78$, respectively. These errors once again indicate that there is less agreement with the observations in terms of the amplitude of the acoustic peaks compared to their multipole positions. Nevertheless, these errors are significantly smaller than those from the Starobinsky inflationary model.

\subsection{Cosmological parameters}

In Table \ref{table2:gStarobinsky}, we compare certain cosmological parameters computed theoretically using the generalized Starobinsky inflationary model with their corresponding observational values derived from the Planck satellite. Once again, all parameters demonstrate alignment between their theoretical predictions and observed values. In fact, all values remain the same as those derived from the Starobisky model ($p=1$), except for the scalar spectral index, which has a smaller uncertainty this time.

\begin{table}[ht!]
\begin{center}
\begin{tabular}{lccc}
\toprule
\textbf{Cosmological Parameter} & \multicolumn{1}{l}{\textbf{Symbol}} & \textbf{G. Starobinsky} & \textbf{Planck}   \\ \midrule
Age of Universe [Gyr]      & Age                             & $13.798 \pm 0.000$      & $13.797 \pm 0.023$   \\
Matter density             & $\Omega_m$                      & $0.3158 \pm 0.0016$     & $0.3153 
\pm 0.073$  \\
Baryon density             & $\Omega_b h^2$                  & $0.02238 \pm  0.0004$   & $0.02237 \pm 0.0001$  \\
Dark energy density                  & $\Omega_{\Lambda}$   & $0.6841 \pm 0.0009$   & $0.6847 \pm 0.0073 $\\
Scalar spectral index               & $n_\sca$                & $0.9672 \pm 0.0024$     & $0.9649 \pm 0.0042$ 
\\ \bottomrule
\end{tabular}
\end{center}
\caption{Cosmological parameters obtained from the generalized Starobinsky inflationary model with $p=1.0004$ compared with the cosmological parameters reported by Planck $2018$ results. }
\label{table2:gStarobinsky}
\end{table}

The percentage errors associated with the age of the universe, matter density, baryon density, dark energy density, and scalar spectral index are $0.007$, $0.16$, $0.04$, $0.09$, and $0.2$, respectively. This model exhibits the same capacity for accurate and precise predictions of the observed cosmological parameters as the Starobinsky inflationary model. However, due to the smaller uncertainty in the scalar spectral index, we can say that this generalized model is slightly better in that aspect, although the difference is not significant.

\section{Chaotic inflationary model with a step}

The chaotic inflationary model with a step was introduce by Adams \cite{adams:2001}, and it is given by

\begin{equation}
\label{V_chaotic}
V(\phi) = \dfrac{1}{2}m^2\phi^{2}\left[1+c\,\tanh\left(\frac{\phi-\phi_\textnormal{step}}{d}\right)\right],   
\end{equation}
where the step occurs at $\phi=\phi_\textnormal{step}$, $\phi$ is the inflaton, $m$ is the inflaton mass, and the parameters $c$ and $d$ are related to the amplitude and width of the feature, respectively \cite{thomas:2023,hernandez:2023,mishra:2020,cadavid:2017,cadavid:2015,adams:2001}.

Using Eqs. \eqref{epsilon}--\eqref{omega} we calculated also, using a Software of symbolic manipulation, the slow--roll parameters for the potential \eqref{V_chaotic} of the chaotic inflationary model with a step, obtaining

\begin{eqnarray}
\label{epsilon_Step}
\epsilon_\nu &=& \dfrac{1}{2 \phi^2} \Bigg[ \dfrac{c \phi F_2^2+2 \left(d+c d F_3 \right)}{ d^2 \left(1+c F_3\right)} \Bigg]^2,\\
\label{eta_Step}
\eta_\nu &=& \dfrac{2}{\phi^2} \Bigg[ 1+\dfrac{c \phi F_2^2 \left(2d-\phi F_3 \right)}{ d^2\left( 1+c F_3\right)} \Bigg],\\
\nonumber
\xi_\nu &=& \dfrac{2}{\phi^3} F_2^2 \left[c\phi F_2^2+2 \left(d+cd F_3 \right) \right]\Bigg[ \dfrac{\left( 3 d^2+2\phi^2-3\phi F_2^2\right) \left(\phi+d F_4\right)}{ d^4\left( 1+c F_3\right)^2} \Bigg],\\
\\
\nonumber
\omega_\nu &=&- \dfrac{8 c}{\phi^4} \left( d-\phi F_3\right)\left(-\phi+3 F_2^2+3dF_3\right) \Bigg[ \dfrac{c\phi F_2^3+2 F_2\left(d+cd F_3 \right)^2}{ d^6\left( 1+c F_3\right)^3} \Bigg],\\
\end{eqnarray} 
where 

\begin{eqnarray}
F_2&=&\textnormal{sech}\left(\dfrac{\phi-\phi_\textnormal{step}}{d}\right),\\
F_3&=&\tanh\left(\dfrac{\phi-\phi_\textnormal{step}}{d}\right), \\
F_4&=&\sinh\left[\dfrac{2\,(\phi-\phi_\textnormal{step})}{d}\right].
\end{eqnarray}

For this model, we need to  calculate numerically the values of $\phi_\text{ini}$ using Eq. \eqref{e-folds} and $\phi_\text{end}$ using Eq. \eqref{epsilon_Step} when $\epsilon_\nu = 1$.
In Fig. \ref{phi_Step}, we show the evolution of the scalar field $\phi(t)$ as a function of time $t$ of the chaotic inflationary model with a step for $c=0.12$,  $d=0.04$, and $\phi_\text{step}=14.09$. The figure illustrates a step in the scalar field because the presence of the feature, and its oscillation after the end of inflation.

\begin{figure}[th!]
\centering
\includegraphics[scale=0.5]{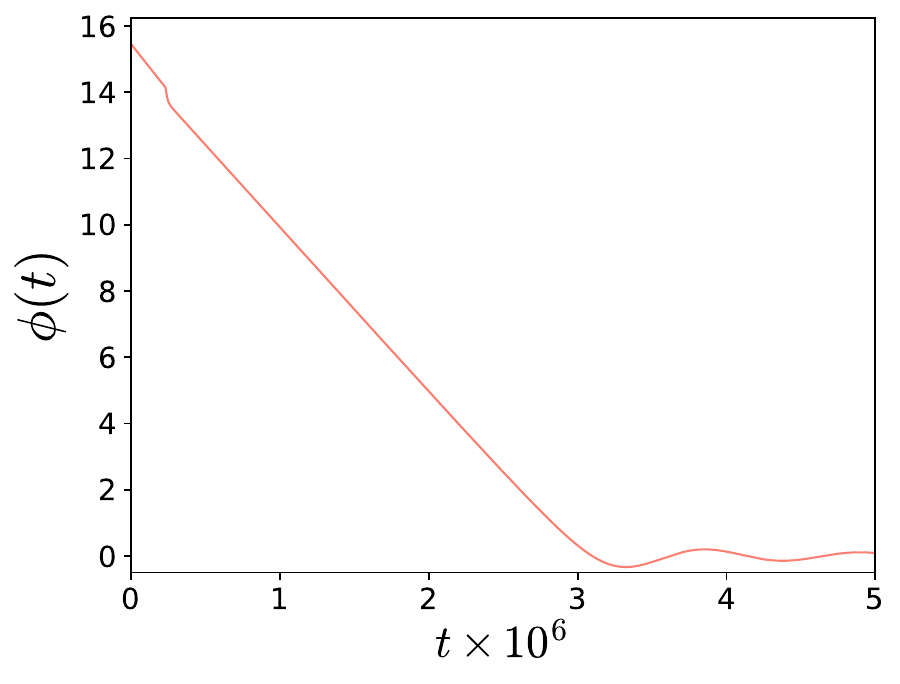}
\caption{Evolution of the scalar field $\phi(t)$ as a function of time $t$ of the chaotic inflationary model with a step for $c=0.12$,  $d=0.04$, and $\phi_\text{step}=14.09$.}
\label{phi_Step}
\end{figure}

Using Eq.\eqref{r} we obtain that the value for the scalar--to--tensor ratio $r$ for the chaotic inflationary model with a step for each considered case is,

\vspace{-0.2cm}
\begin{table}[ht!]
\begin{center}
\begin{tabular}{cccc}
\toprule
$c$ & $d$ & $\phi_\textnormal{step}$  & $r$\\
\midrule
0 & 0.04 & 14.09 & 0.132232\\
0.0012 & 0.04 & 14.09 & 0.132232\\
0.012 & 0.04  & 14.09 & 0.132232\\
0.12 & 0.04  & 14.09 & 0.132232\\
\bottomrule
\end{tabular}
\end{center}
\caption{Value of the scalar--to--tensor ratio $r$ for each value of $c$,  $d$, and $\phi_\textnormal{step}$ considered in the chaotic inflationary model with a step.}
\label{r:Step}
\end{table}

Fig. \ref{Step_epsilon} and \ref{Step_eta} show   the slow--roll parameters \eqref{epsilon_Step} and \eqref{eta_Step} for the chaotic inflationary model with a step, in which we can observed that slow--roll parameters are violet at $N_e \sim 10$ just where the step occurs, it is because the presence of the feature \cite{hernandez:2023,mishra:2020}.

\begin{figure}[th!]
\begin{subfigure}[b]{0.5\textwidth}
\centering
\includegraphics[width=0.96\textwidth]{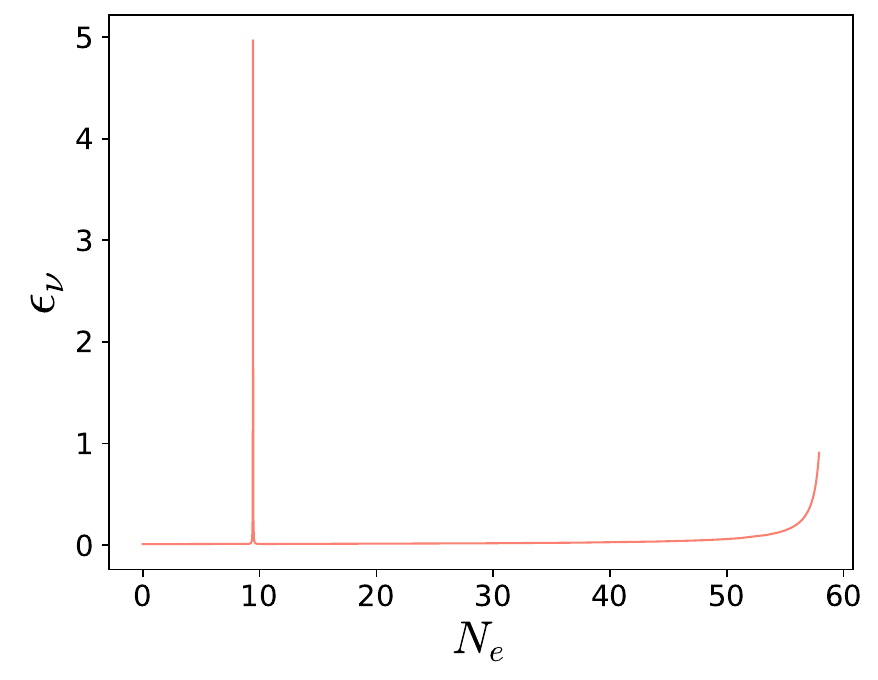}
\caption{}
\label{Step_epsilon}
\end{subfigure}
\begin{subfigure}[b]{0.5\textwidth}
\centering
\includegraphics[width=\textwidth]{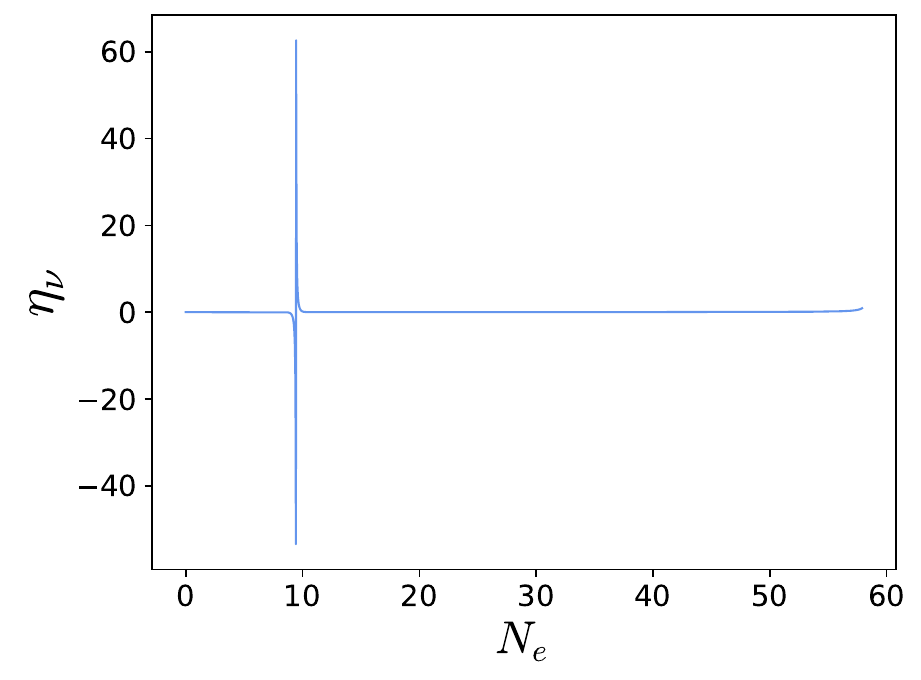}
\caption{}
\label{Step_eta}
\end{subfigure}
\caption{Behaviour of the slow--roll parameters $\epsilon_\nu$ and $\eta_\nu$ with respect to the number of e--folds $N_e$, described by the chaotic inflationary model with a step inflationary model for $c=0.12$,  $d=0.04$, and $\phi_\text{step}=14.09$.}
\end{figure}

Fig. \ref{Step_1} shows the CMB angular power spectrum for the chaotic inflationary model with a step, different values of the parameters $c$ and $d$ were used to reproduce different spectra using \texttt{CAMB}. In Fig. \ref{Step_2} we report the best fit compared to Planck $2018$ data, for this fit we use $d = 0.04$, $c = 0.12$, and $\phi_\text{step}=14.09$. Our best fit was chosen by computing the relative error between our model and the Planck $2018$ data and selecting the parameters that produce the lower relative error. We can see from Fig. \ref{error_Step_2} that the relative error found for these parameters is lower than 9\% regarding the spectrum from Planck $2018$ data.

\begin{figure}[th!]
\centering
\includegraphics[width=\textwidth]{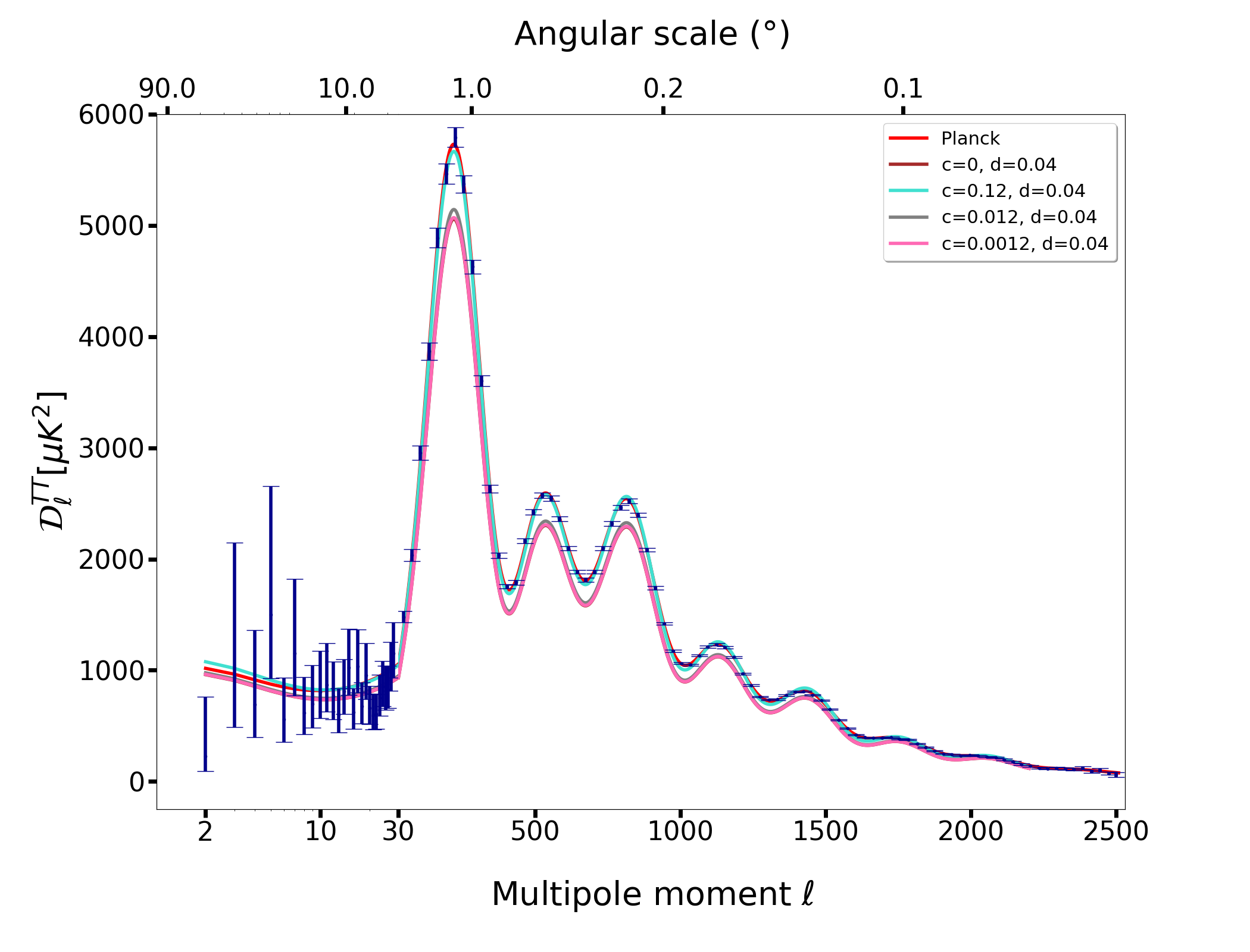}
\caption{Angular power spectrum for the chaotic inflationary model with a step  reproduced by \texttt{CAMB} for different values of $c$ and $d$. Red   line: Planck $2018$ data, brown line:  result for $c=0$, turquoise line:  result for $c=0.12$ and $d=0.04$, gray line: result for $c=0.012$ and $d=0.04$, and pink line: result for $c=0.0012$ and $d=0.04$.}
\label{Step_1}
\end{figure}

In Fig. \ref{Step_3} is observed the large angular scales ($\ell < 90$) at $c=0.12$ and $d=0.04$, and its relative error is shown in Fig. \ref{error_Step_3}. 
In Fig. \ref{Step_4} is observed the intermediate angular scales ($90 < \ell \lesssim 900$) at $c=0.12$ and $d=0.04$, and its relative error is shown in Fig. \ref{error_Step_4}.
In Fig. \ref{Step_5} is observed the large angular scales ($\ell > 900$) at $c=0.12$ and $d=0.04$, and its relative error is shown in Fig. \ref{error_Step_5}.

This model reproduces a CMB that aligns better with the Planck $2018$ observations, surpassing the Starobinsky and generalized Starobinsky inflationary models. High--variability in the results persists at small angular scales as in the previous models. To assess the overall balance of these errors, let us examine the L$2$ norm.

When considering all scales, this inflationary model has a L$2$ norm error of 26 $\mu K^2$. On large, intermediate and small angular scales, the L2 norm errors are 20 $\mu K^2$, 36 $\mu K^2$, and 18 $\mu K^2$, respectively. Therefore, the chaotic model exhibits similar characteristics as the models seen until now. It reproduced the CMB angular spectrum more accurately at smaller angular scales, but the error balance is less favorable at intermediate scales. But, in general, it is the best model for reproducing the CMB angular spectrum,  both in terms of relative error size and error balance.

\begin{figure}[th!]
\centering
\begin{subfigure}[b]{0.50\textwidth}
\centering
\includegraphics[width=\textwidth]{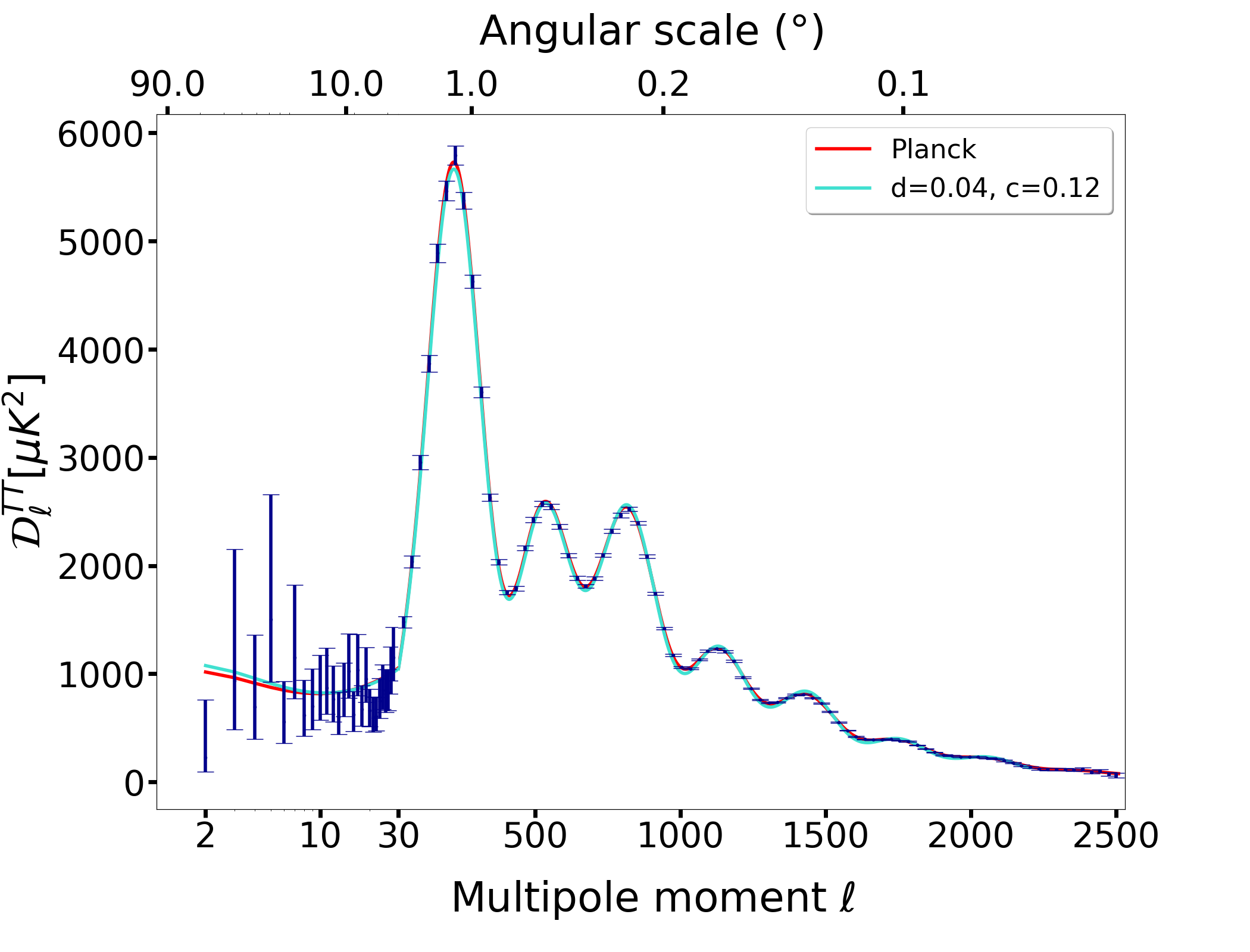}
\caption{}
\label{Step_2}
\end{subfigure}
\begin{subfigure}[b]{0.49\textwidth}
\centering
\includegraphics[width=\textwidth]{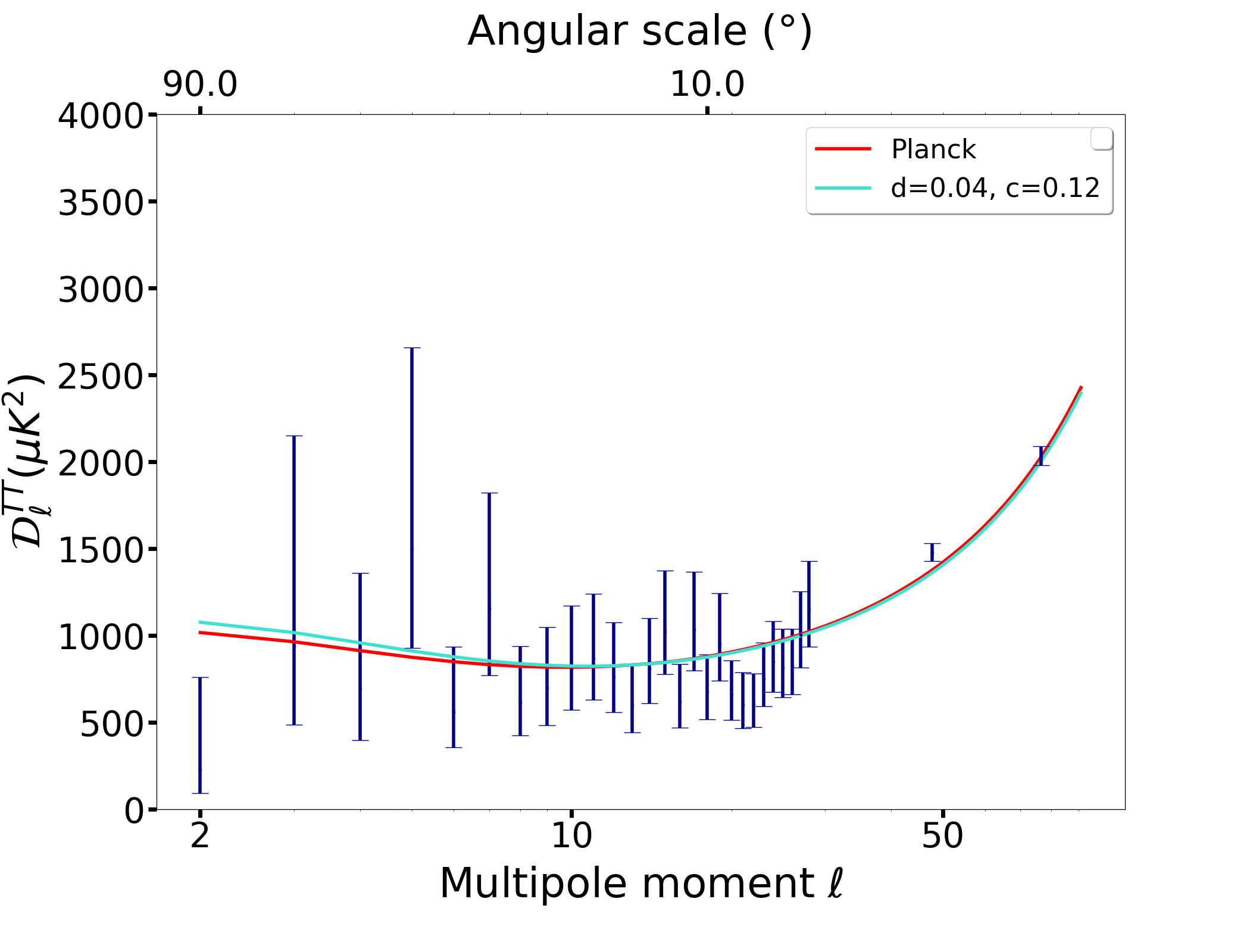}
\caption{}
\label{Step_3}
\centering
\end{subfigure}
\begin{subfigure}[c]{0.49\textwidth}
\includegraphics[width=\textwidth]{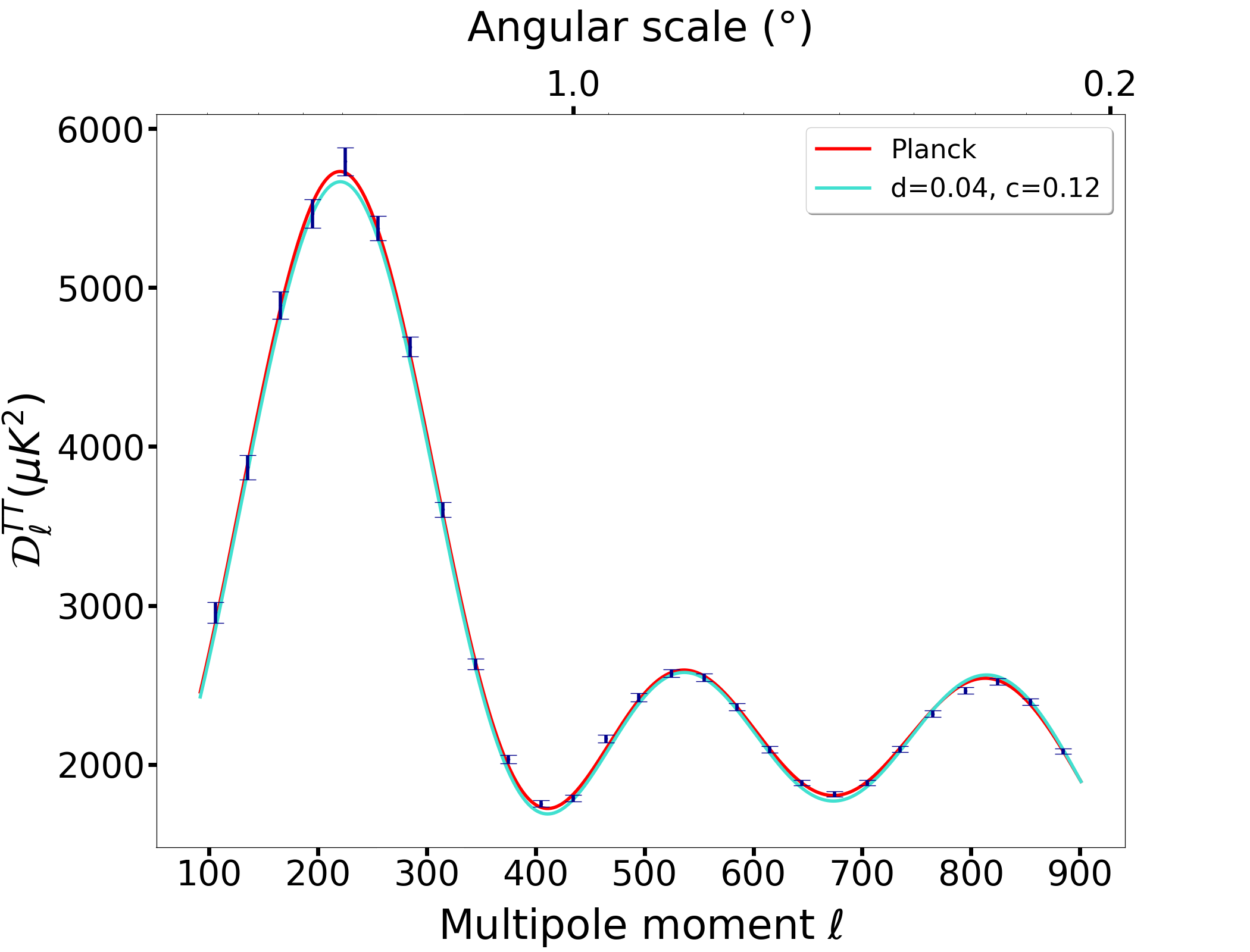}
\caption{}
\label{Step_4}
\end{subfigure}
\begin{subfigure}[d]{0.49\textwidth}
\centering
\includegraphics[width=\textwidth]{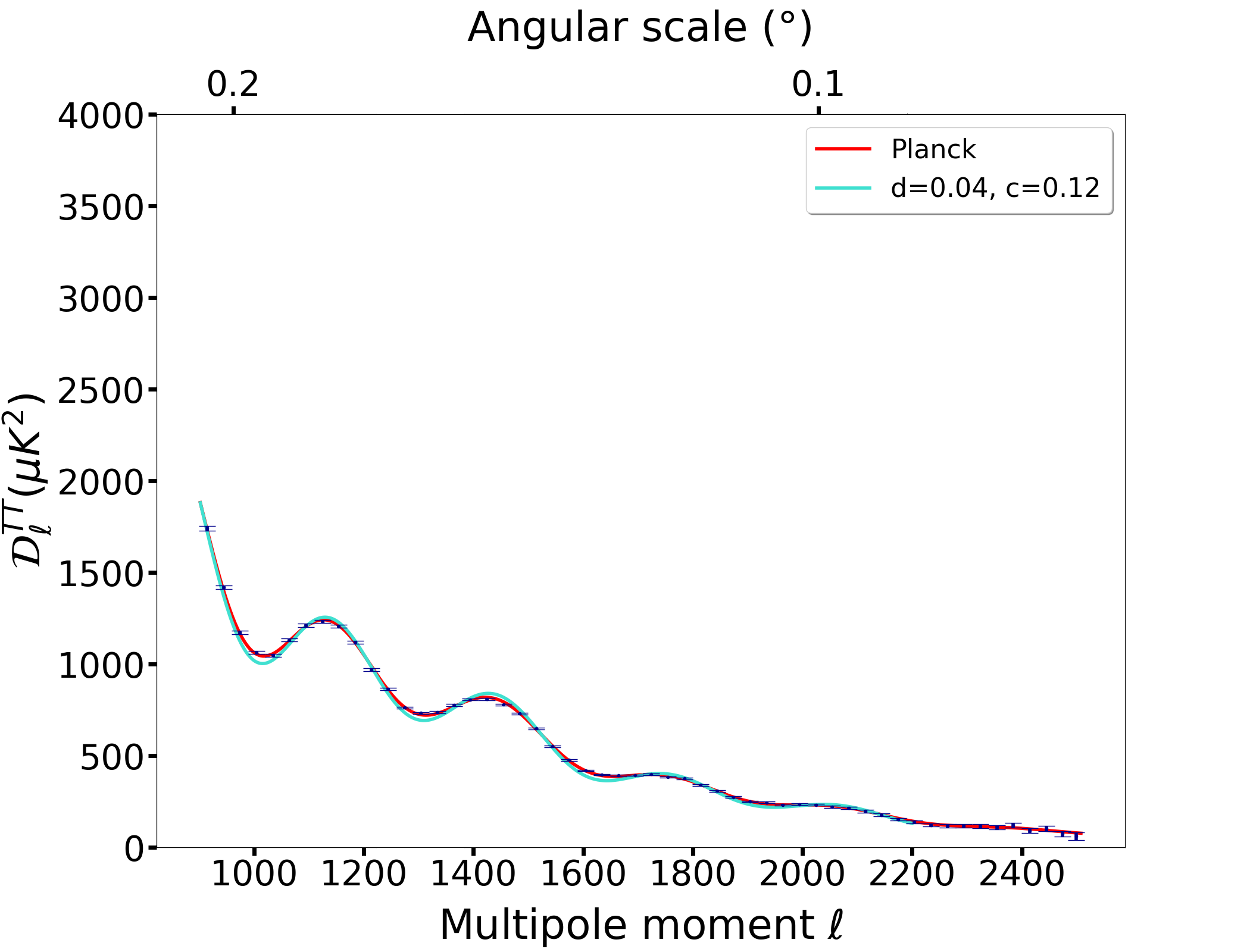}
\caption{}
\label{Step_5}
\end{subfigure}
\caption{(a) Angular power spectrum for the chaotic inflationary model with a step  reproduced by \texttt{CAMB}  for  $c=0.12$ and $d=0.04$, (b) large angular scales, (c) intermediate angular scales, and (d) small angular scales. Red  line: Planck $2018$ data with error bars in blue, turquoise line: \texttt{CAMB} result.}
\end{figure}

\begin{figure}[th!]
\centering
\begin{subfigure}[b]{0.50\textwidth}
\centering
\includegraphics[width=\textwidth]{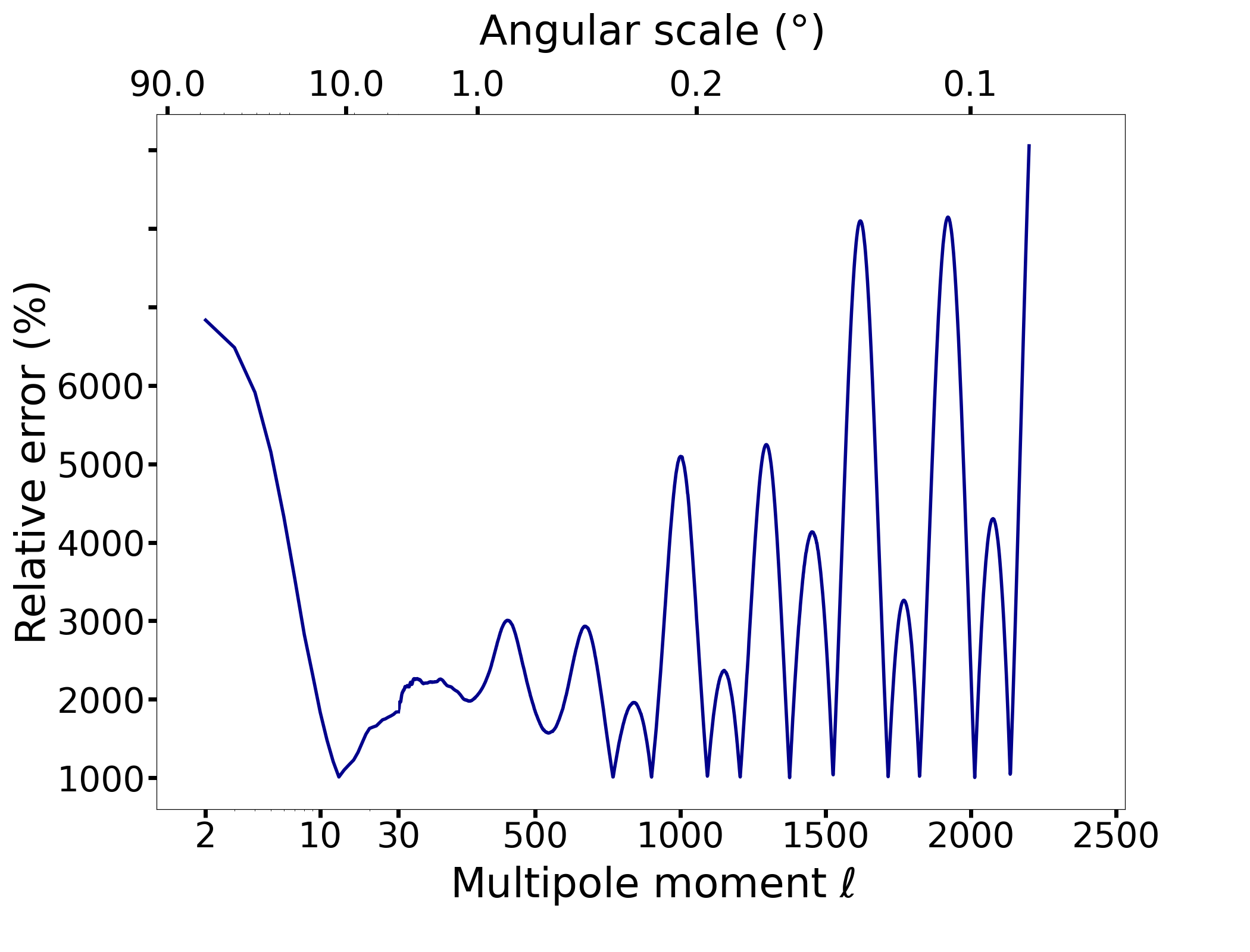}
\caption{}
\label{error_Step_2}
\end{subfigure}
\begin{subfigure}[b]{0.49\textwidth}
\centering
\includegraphics[width=\textwidth]{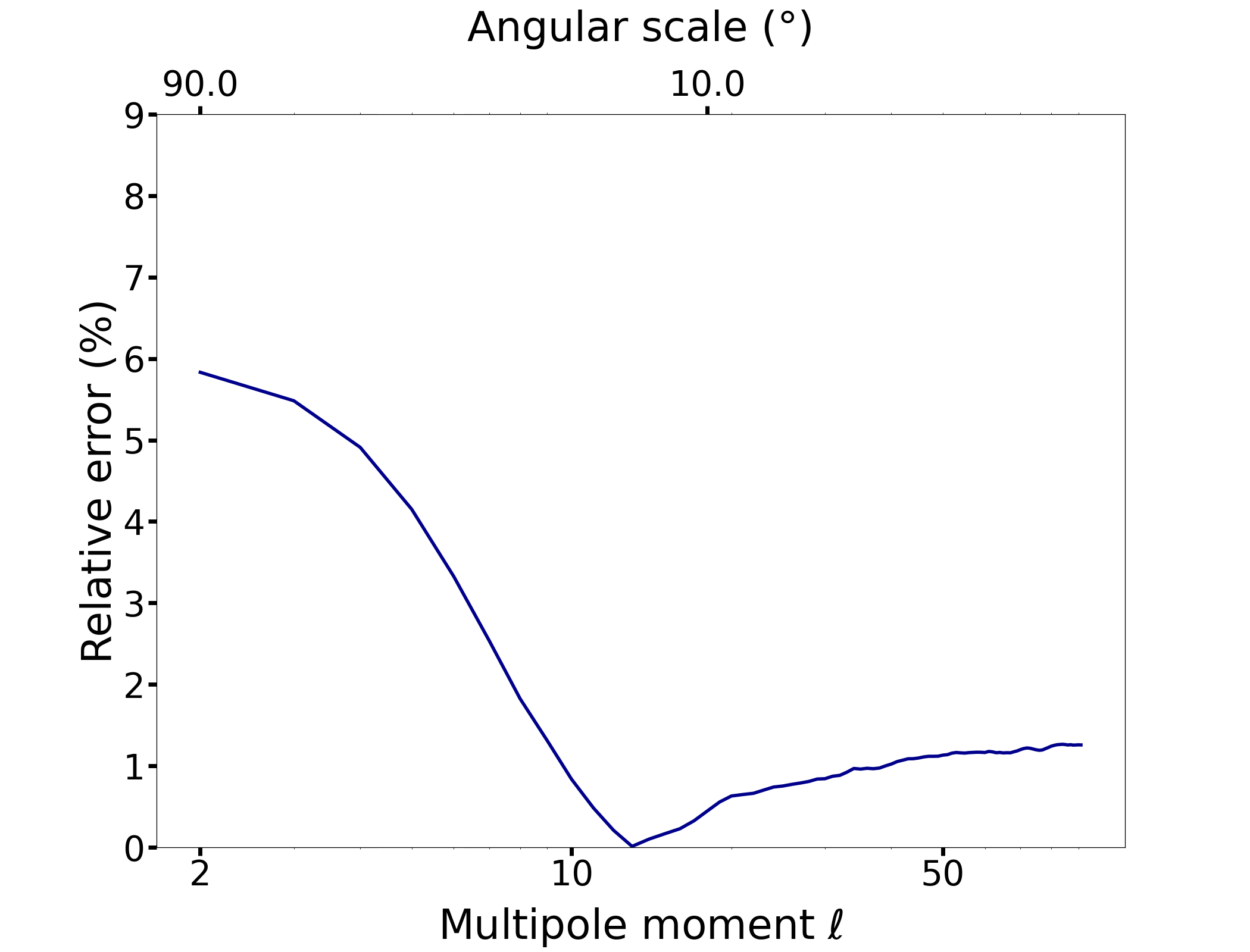}
\caption{}
\label{error_Step_3}
\centering
\end{subfigure}
\begin{subfigure}[c]{0.49\textwidth}
\includegraphics[width=\textwidth]{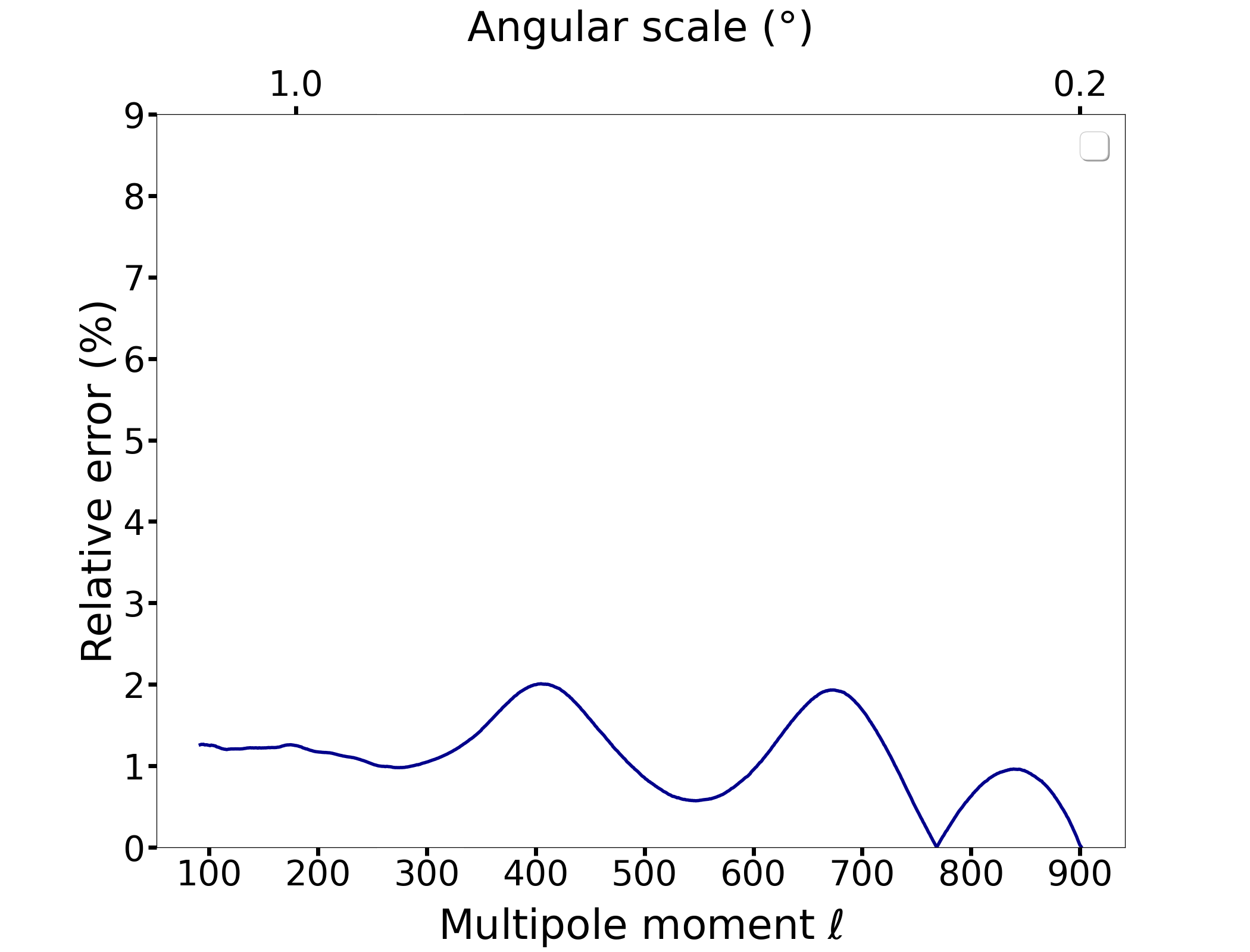}
\caption{}
\label{error_Step_4}
\end{subfigure}
\begin{subfigure}[d]{0.49\textwidth}
\centering
\includegraphics[width=\textwidth]{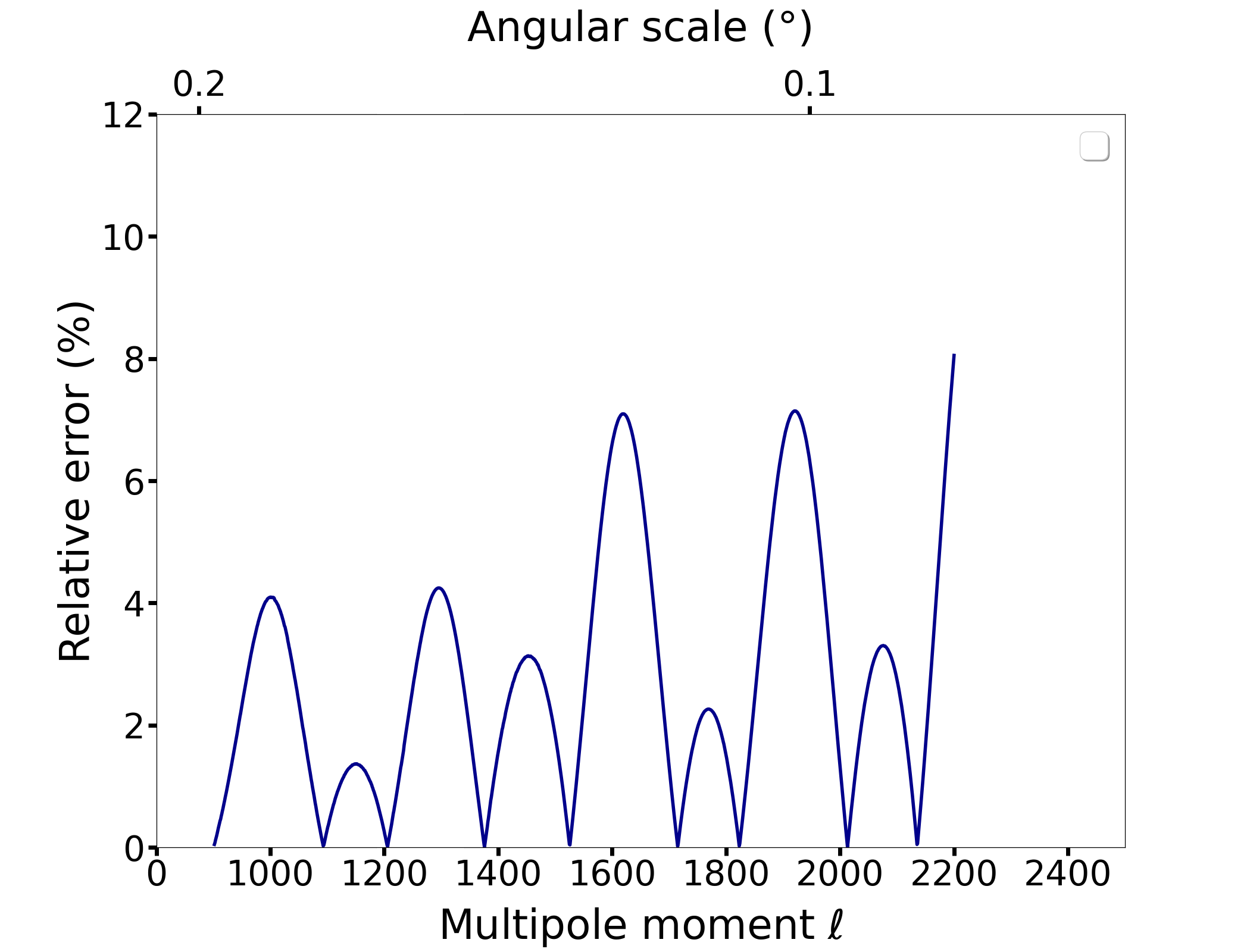}
\caption{}
\label{error_Step_5}
\end{subfigure}
\caption{(a) Error of the angular power spectrum for the chaotic inflationary model with a step reproduced by \texttt{CAMB}, (b) large angular scales, (c) intermediate angular scales, and (d) small angular scales.}
\end{figure}

\subsection{The Acoustic Peaks}

In Table \ref{table1:Step}, we present the acoustic peaks identified in the intermediate scale of the spectrum, together with the corresponding troughs in this range.

\begin{table}[th!]
\begin{center}
\makebox[\linewidth]{
\begin{tabular}{lcccc}
\toprule
& \multicolumn{2}{c}{\textbf{Inflationary chaotic model with a step}} & \multicolumn{2}{c}{\textbf{Planck Satellite}}\\
\textbf{Extreme} & \textbf{Multipole [$\ell$]} & \textbf{Amplitude [$\mu K^2$]} & \textbf{Multipole [$\ell$]} & \textbf{Amplitude [$\mu K^2$]}\\ 
\midrule
Peak $1$   & 221 & 5143.42 $\pm$ 0.10 & 220.6 $\pm$ 0.6 & 5733 $\pm$ 39 \\
Trough $1$ & 411 & 1532.55 $\pm$ 0.11 & 416.3 $\pm$ 1.1 & 1713 $\pm$ 20 \\
Peak $2$   & 537 & 2340.94 $\pm$ 0.10 & 538.1 $\pm$ 1.3 & 2586 $\pm$ 23 \\
Trough $2$ & 674 & 1606.23 $\pm$ 0.11 & 675.5 $\pm$ 1.2 & 1799 $\pm$ 14 \\
Peak $3$   & 814 & 2326.50 $\pm$ 0.08 & 809.8 $\pm$ 1.0 & 2518 $\pm$ 17 \\
\bottomrule
\end{tabular}
}
\end{center}
\caption{Peaks and troughs of the CMB  TT  power spectra in the Acoustic Peak region recreated by the chaotic inflationary model with a step for $p=1.0004$ and reported by Planck satellite.}
\label{table1:Step}
\end{table}

Once again, our observations reveal the proximity between the positions on the multipole axis of both observed and computed peaks. Taking into account the errors associated with the observed values, all computed peaks, except the third one, align well with the data from the Planck satellite. Specifically, the percentage errors for the calculated multipole values corresponding to peak $1$, trough $1$, peak $2$, trough $2$, and peak $3$ are $0.18$, $1.27$, $0.20$, $0.22$, and $0.52$, respectively. We can see again the same percentage error of the previously presented models.

The percentage errors for the calculated amplitudes of peak $1$, trough $1$, peak $2$, trough $2$, and peak $3$ are $10.28$, $10.53$, $9.48$, $10.72$, and $7.61$, respectively. Once again, there is less agreement between the observations and the computed values in terms of the amplitude of the acoustic peaks compared to their multipole positions. Although the computed data generally fits better with the observations than the previous models, it exhibits greater percentage errors when comparing the acoustic peaks in the intermediate angular scale of the spectrum.

\subsection{Cosmological parameters}

In Table \ref{table2:Step}, we once again analyze the same cosmological parameters derived theoretically from the chaotic model and compare them with their corresponding observational values obtained from the Planck satellite. Considering uncertainties, all the parameters exhibit alignment between their theoretical and observational values.

\begin{table}[th!]
\begin{center}
\begin{tabular}{lccc}
\toprule
\textbf{Cosmological Parameter} & \multicolumn{1}{l}{\textbf{Symbol}} & \textbf{Chaotic} & \textbf{Planck}   \\ \midrule
Age of Universe [Gyr]      & Age                             & $13.798 \pm 0.000$      & $13.797 \pm 0.023$   \\
Matter density             & $\Omega_m$                      & $0.3158 \pm 0.0016$     & $0.3153 
\pm 0.073$  \\
Baryon density             & $\Omega_b h^2$                  & $0.02238 \pm  0.0004$   & $0.02237 \pm 0.0001$  \\
Dark energy density                  & $\Omega_{\Lambda}$   & $0.6841 \pm 0.0009$   & $0.6847 \pm 0.0073 $\\
Scalar spectral index               & $n_\sca$               & $0.9669 \pm 0.0021$     & $0.9649 \pm 0.0042$ 
\\ \bottomrule
\end{tabular}
\end{center}
\caption{Cosmological parameters obtained from the chaotic inflationary model with a step for $p=1.0004$ compared with the cosmological parameters reported by Planck $2018$ results.}
\label{table2:Step}
\end{table}

The percentage errors for the age of the universe, matter density, baryon density, dark energy density, and scalar spectral index are $0.007$, $0.16$, $0.04$, $0.09$, and $0.2$, respectively. The effectiveness, accuracy, and precision demonstrated by the chaotic model are again similar to those of the previous ones. It can also be stated that it is slightly better, mainly due to the better precision it achieves when predicting the scalar spectral index.

\newpage
\section{Analysis of the Model Comparison}

To assess the validity of different inflationary models with respect to the observed Cosmic Microwave Background (CMB) data, we calculated the \(\chi^2\) statistic for each model, this was inspired in the analysis made for Ade \textit{et al.} \cite{ade:2016}. The \(\chi^2\) statistic is given by:

\begin{equation}
\chi^2 = \sum_{i=1}^{n} \frac{(M_i - O_i)^2}{\sigma_i^2},
\end{equation}
where \(M_i\) represents the model prediction, \(O_i\) is the observed data, and \(\sigma_i\) is the associated uncertainty. To provide a fair comparison across models, we normalized the \(\chi^2\) by the degrees of freedom \(\nu\), which is defined as:

\begin{equation}
\nu = N_{\text{data}} - N_{\text{parameters}}.
\end{equation}

This normalization yields the reduced \(\chi^2\):

\begin{equation}
\chi^2_{\nu} = \frac{\chi^2}{\nu}.
\end{equation}

The results for the three inflationary models under consideration are as follows in Table \ref{table1:chi2}.

\begin{table}[th!]
\begin{center}
\makebox[\linewidth]{
\begin{tabular}{lc}
\toprule
Inflationary Model                         & $\chi_\nu^2$ \\
\midrule
Starobinsky inflationary model             & 7.52 \\
Generalized Starobinsky inflationary model & 0.22 \\
Chaotic inflationary model with a step     & 0.15 \\
\bottomrule
\end{tabular}
}
\end{center}
\caption{$\chi_\nu^2$ calculated by Starobinsky inflationay model, the generalized Starobinsky inflationary model for $p=1.0004$, the chaotic inflationary model with a step using $c=012$ and $d=0.04$.}
\label{table1:chi2}
\end{table}

The \(\chi^2_{\nu}\) value for the Starobinsky inflationary model, 7.52, indicates a poor fit to the CMB data, as a \(\chi^2_{\nu}\) significantly greater than 1 suggests that the model overpredicts the observed values or does not capture the underlying physics accurately. This result implies that the standard Starobinsky model may require further modifications or that it might not be suitable under the current observational constraints.

In contrast, the generalized Starobinsky inflationary model yields a \(\chi^2_{\nu} = 0.22\), and the chaotic inflationary model with a step results in \(\chi^2_{\nu} = 0.15\). Both values are significantly lower than 1, suggesting an overfitting scenario where the models are too flexible relative to the data. These low \(\chi^2_{\nu}\) values could imply that while these models fit the data well, they may do so at the cost of excessive complexity, capturing noise or fluctuations in the observational data rather than genuine physical features.

Between the two models with low \(\chi^2_{\nu}\), the chaotic inflationary model  with a step exhibits the lowest value, which might be interpreted as slightly better in terms of fitting the observed data. However, such a low \(\chi^2_{\nu}\) value must be approached with caution, as it may also reflect a lack of robustness rather than true physical relevance.

\section{Bayesian Analysis}

In order to made a selection of our three best inflationary models also we have perform a Bayesian Analysis using the \texttt{Cobaya} code \cite{torrado:2021} varying the effective number of neutrino species  $N_\text{eff}$ an extra $\Delta N_\text{eff}$.

After calculate the Bayesian evidence $-\text{log}(Z)$ \cite{tram:2017} using the algorithm \texttt{PolyChord} \cite{handley:2015} we found the following results: 

\begin{table}[th!]
\begin{center}
\makebox[\linewidth]{
\begin{tabular}{lc}
\toprule
Inflationary Model                         & $-\text{log}(Z)$ \\
\midrule
Starobinsky inflationary model             &$ 0.803\times 10^{5}\pm 0.384\times10^{1} $ \\
Generalized Starobinsky inflationary model & $ 0.725\times 10^{5}\pm 0.376\times10^{1} $ \\
Chaotic inflationary model with a step     &$ 0.764\times 10^{5}\pm 0.377\times10^{1} $ \\
\bottomrule
\end{tabular}
}
\end{center}
\caption{The evidence $Z$ is a log--normally distributed, with location and scale parameters $\mu$ and $\sigma$. We denote this as $-\text{log}(Z) = \mu \pm \sigma$.}
\label{evidence}
\end{table}

From table \ref{evidence}, it is necessary to calculate the Bayes ratios $B_{ij}=-\text{log}\left(\frac{Z_i}{Z_j}\right)$, which are essential to model selection processes \cite{keeley2022distribution}. In order to facilitate the calculations let us denote $Z_1$, $Z_2$, $Z_3$ as the Bayesian evidences for the Starobinsky inflationary model, generalized Starobinsky inflationary model, and chaotic inflationary model with a step, respectively.

\subsection{Starobinsky inflationary model vs. generalized Starobinsky inflationary model}

\begin{equation}
B_{21}=\dfrac{0.725}{0.803}\approx 0.903.
\label{b1}
\end{equation}

This means the generalized Starobinsky model is about 0.903 times as likely as the Starobinsky inflationary model.

\subsection{Chaotic inflationary model with a step model vs. generalized Starobinsky inflationary model}
\begin{equation}
B_{23}=\dfrac{0.725}{0.764}\approx 0.949.
\label{b2}
\end{equation}

This means the generalized Starobinsky inflationary model is about 0.949 times as likely as the chaotic inflationary model with a step.

\subsection{Starobinsky inflationary model vs. chaotic  inflationary model with a step}
\begin{equation}
B_{31}=\dfrac{0.764}{0.803}\approx 0.951.
\label{b3}
\end{equation}

This means the chaotic inflationary model with a step is about $0.951$ times as likely as the Starobinsky inflationary model.

Analyzing the results above it is possible to conclude that generalized Starobinsky and the chaotic with a step inflationary models are better that the Starobinsky inflationary model. It happens because the Bayes factor corresponds to the relative probability of one model over another  \cite{keeley2022distribution}. 

\begin{figure}[th!]
\includegraphics[width=\textwidth]{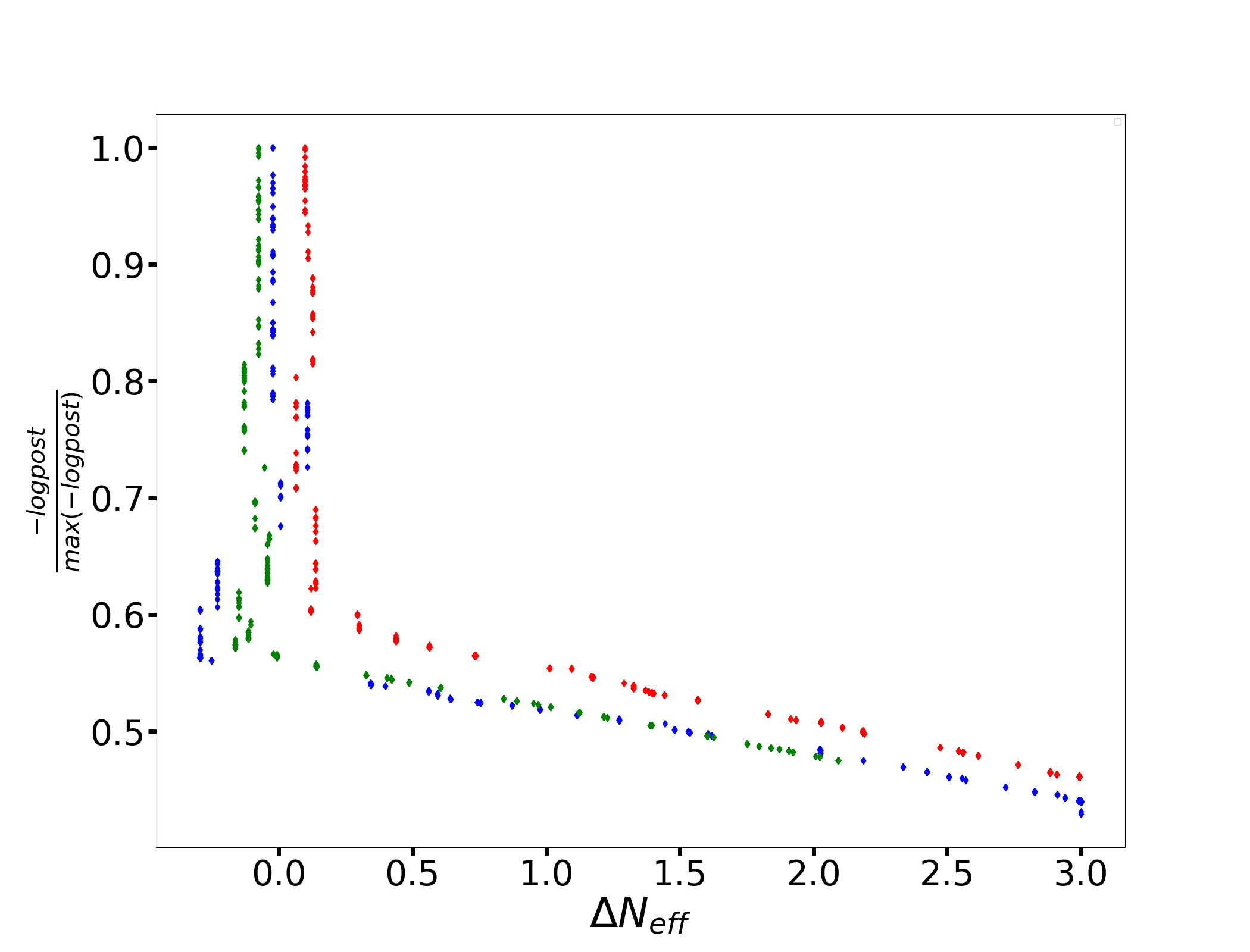}
\caption{Posterior distributions on the effective number of neutrino species $\Delta N_\text{eff} = N_\text{eff}-3.046$
for the single--field models discussed in this paper. Starobinsky model inflationary model (red--dotted line), chaotic inflationary model with a step model (green--dotted line), and generalized Starobinsky inflationary model (blue--dotted line).}
\label{posterior}
\end{figure}

The previous fact is corroborated by the data presented in Figure \ref{posterior}, where we show the negative logarithm of the posterior distributions of $\Delta N_\text{eff}$. It is clear to see that the posteriors of the generalized Starobinsky and chaotic with a step models are similar. Moreover, there is a good amount of data that is near  to $0$ which means that $\Delta N_\text{eff}\approx 0$ or $N_\text{eff}\approx 3.046$. The posterior distribution for the Starobinsky model differs from the other distributions which means that other models are more accurate than the Starobinsky model.

\section{Constrain of our inflationary models from reheating}

In this section we perform an analysis of all our inflationary models considering the number of e-foldings from the end of inflation until the end of reheating, which is given by  \cite{cook:2015reh},

\begin{equation}
N_{\text{reh}}=\dfrac{4}{1-3 \bar{\omega}_{\text{reh}}} \left[61.6-\ln\left(\dfrac{V_{\text{end}}^{\sfrac{1}{4}}}{H_k} \right)- N_k \right],
\label{Nreh}
\end{equation}
where $ \bar{\omega}_{\text{reh}}$ is the averaged equation of state during
reheating,   $V_{\text{end}}$ is the value of the potential at the end of inflation, $H_k$ is the Hubble parameter at the Planck's pivot scale $k_*$, and  $N_k$ is the number of e--foldings between the exit time of the modes at this pivot during inflation and the end of inflation, which is given by Eq. \eqref{e-folds} evaluated at $\phi=\phi_*$. Eq. \eqref{Nreh} is a simplified version of the complete equation given in reference \cite{drewes:2017reh} considering the number of degrees of freedom of species at the end of reheating  equal to the entropy number of degrees of freedom after reheating $g_\text{reh}=g_{\text{s,reh}} \approx 100$, the amplitude of the scalar power spectrum $A_\sca= 2.1 \times 10^{-9}$, the Planck's pivot scale $k_*$ is equal to $0.05$ Mpc$^{-1}=1.31 \times 10^{-58}$ and, the current temperature of the CMB as $T_0=2.725 \,\text{K}=9.62 \times 10^{-32}$   
 \cite{german:2022reh}.

Following the procedure of Martin \textit{et al.} \cite{martin:2010reh, martin:2015reh}, and Cook \cite{cook:2015reh},  we analyze the dependence of our observables $n_\sca$ and $r$ by varying the number of e-folds during reheating $N_{\text{reh}}$ with $\bar{\omega}_{\text{reh}}$ for each inflationary model. Then we evaluate these quantities at the duration of reheating $\Delta N \equiv  N_{\text{reh}}-N_e $.  The duration of the reheating period affects the relationship between the scalar spectral index $n_\sca$ and the scalar--to--tensor ration $r$ and producing different contours.

\subsection{Starobinsky inflationary model}

\bigskip
For the Starobinsky inflationary model the scalar spectral index $n_\sca$ and the tensor--to--scalar ratio $r$  in terms of the number of e--folds are calculated from Eqs. \eqref{nS} and \eqref{r} using the slow--roll parameters, and are given by \cite{renzi:2020},

\begin{eqnarray}
n_\sca&\simeq&1-\dfrac{2}{\Delta N},\\
r&\simeq&\dfrac{12}{\Delta N^{2}}.
\end{eqnarray}

\bigskip
Fig. \ref{contour_Starobinsky} shows the contour plot $(n_\sca,r)$ for the Starobinsky inflationary model for three values of $\bar{\omega}_{\text{reh}}=-0.3, -0.2, 0$ where we can observed  the influence of  reheating in our observables.

\begin{figure}[th!]
\includegraphics[width=\textwidth]{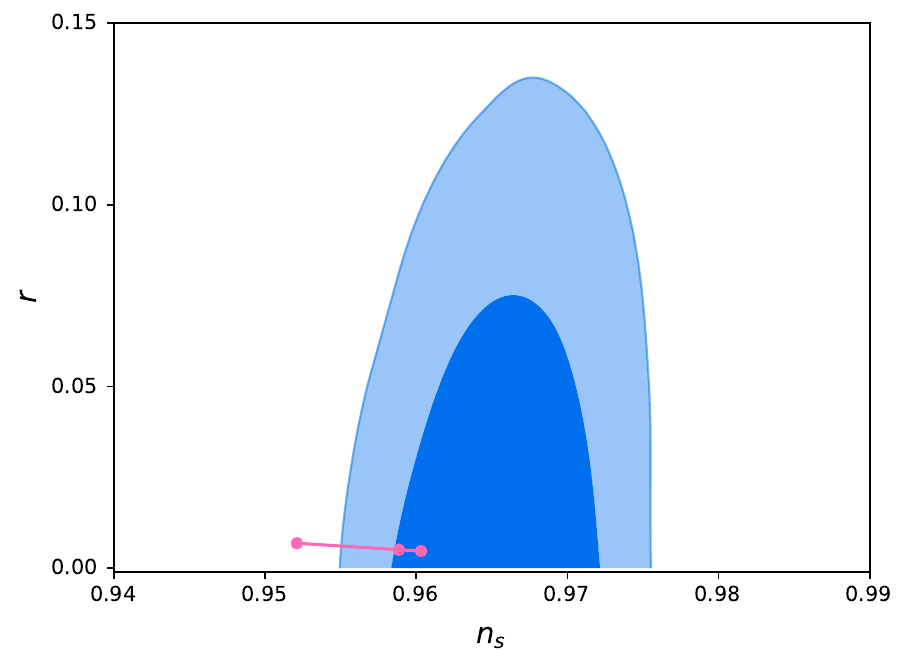}
\caption{Contour plot $(n_\sca,r)$ for the Starobinsky inflationay model.}
\label{contour_Starobinsky}
\end{figure}
 
 \newpage
\subsection{Generalized Starobinsky inflationary model}

\bigskip
For the generalized Starobinsky inflationary model the  scalar spectral index $n_\sca$ and the tensor--to--scalar ratio $r$ in terms of the number of e--folds from the slow--roll parameters, are given by \cite{renzi:2020},

\begin{eqnarray}
n_\sca&\simeq&1-\dfrac{8(p-1)\left[C^{2}(p-1)-p(C-1)\right]}{3\left[C(1-2p)+p\right]^2},\\
r&\simeq&\dfrac{64 C^2(p-1)^2}{3\left[C(1-2p)+p\right]^2},
\end{eqnarray}
where $C \equiv  e^{-\frac{4}{3}\Delta N (p-1)}$.

\bigskip
Fig. \ref{contour_gStarobinsky} shows the contour plot $(n_\sca,r)$ for the generalized Starobinsky inflationary model considering three different values of $\bar{\omega}_{\text{reh}}=-0.3, -0.2, 0$, and twelve different values for the parameter $p$. In this plot  we can observed  the influence of the reheating in our observables, and how the generalized Starobinsky inflationary model with $0.995 \leq p \leq 1.0004$ reaches the $95\%$ confidence region based on Planck 2018 data when  the averaged equation of state during reheating is set to $\bar{\omega}_{\text{reh}}=-0.3$.

\begin{figure}[th!]
\includegraphics[width=\textwidth]{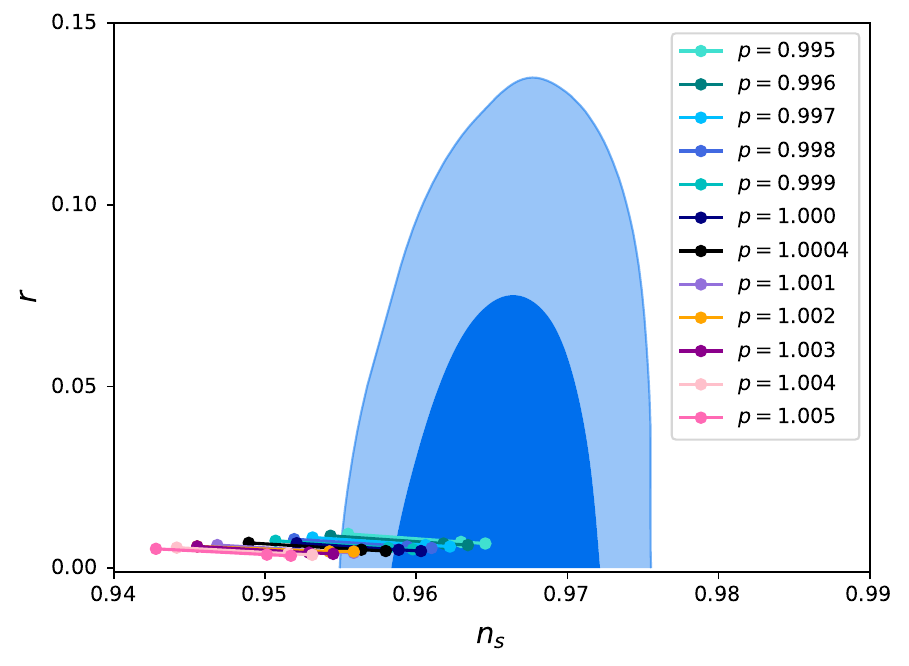}
\caption{Contour plot $(n_\sca,r)$ for the generalized Starobinsky inflationary model for various values of the parameter $p$.}
\label{contour_gStarobinsky}
\end{figure}

\subsection{Chaotic inflationary model with a step}

\bigskip
For the chaotic inflationary model with a step the scalar spectral index $n_\sca$ and the tensor--to--scalar ratio $r$ in terms of the number of e--folds are given by \cite{dimarco:2024reh},

\begin{eqnarray}
n_\sca&\simeq&1-\dfrac{2}{\Delta N},\\
r&\simeq&\dfrac{8}{\Delta N}.
\end{eqnarray}

\bigskip
Fig. \ref{contour_Step} shows the contour plot $(n_\sca,r)$ for the generalized chaotic inflationary model with a step considering three different values of $\bar{\omega}_{\text{reh}}=-0.3, -0.2, 0$, and  four set of values of the parameters $c$ and $d$. In this plot  we can observed  that regardless of the three values of $\bar{\omega}_{\text{reh}}$ , no model enters the $63\%$ confidence region of the Planck 2018 data.

\begin{figure}[th!]
\includegraphics[width=\textwidth]{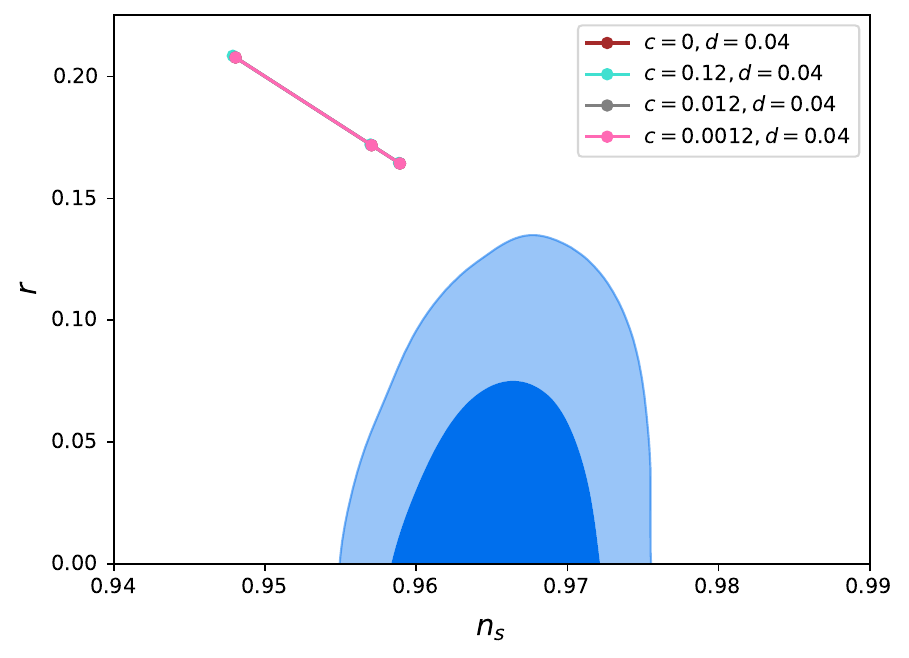}
\caption{Contour plot of $(n_\sca,r)$ for the chaotic inflationary model with a step for various values of the parameters $c$ and $d$.}
\label{contour_Step}
\end{figure}

\section{Conclusions}

In summary, this article explores and analyzes three different inflationary models: the Starobinsky inflationary model, the generalized Starobinsky inflationary model, and the chaotic inflationary model with a step. 

For the Starobinsky inflationary model, the inflationary potential is given by Eq. \eqref{V_Starobinsky}, and the slow--roll parameters are calculated analytically. The CMB angular power spectrum for this model is presented and compared with Planck $2018$ data in various plots, showing good agreement at small angular scales but with larger errors at intermediate scales. The acoustic peaks and troughs are also listed, and cosmological parameters calculated from the model are compared with observational values, demonstrating overall agreement.

The generalized Starobinsky inflationary model, described by Eq. \eqref{V_gStarobinsky}, introduces an additional parameter $p$, and the slow--roll parameters are derived. The CMB angular power spectrum for various values of $p$ is presented, and the optimal value that best fits the Planck $2018$ data is determined to be $p=1.0004 \pm 0.0001$. The model exhibits improved performance compared to the original Starobinsky inflationary model, with smaller errors in the relative error plots. The acoustic peaks and cosmological parameters are also analyzed, showing good agreement with observations.

While the generalized Starobinsky and chaotic models provide a much better fit to the CMB data than the original Starobinsky model, their low \(\chi^2_{\nu}\) values suggest potential overfitting. Further investigations are necessary to evaluate whether these models' complexity can be justified or if additional constraints are needed to prevent overfitting. Future work will focus on refining these models and exploring alternative inflationary scenarios that strike a balance between fit quality and model simplicity.

Regarding the reheating constraints of our inflationary models, the generalized Starobinsky inflationary model with $0.995 \leq p \leq 1.0004$  and $\bar{\omega}_{\text{reh}}=-0.3$ offers a better positioning within the $(n_\sca,r)$ contour plot.

Finally, the chaotic inflationary model with a step, described by Eq. \eqref{V_chaotic}, is investigated. The slow--roll parameters are derived, and the CMB angular power spectrum is presented. The model introduces a step in the potential and its impact on the power spectrum is analyzed. The model is characterized by parameters such as $m$, $c$, $d$, and $\phi_\textnormal{step}$.  

Overall, all three inflationary models successfully replicate the shape of the temperature power spectrum of the CMB while aligning with the values of cosmological parameters. The chaotic inflationary model with a step produce better results in the positioning and amplitudes of the peaks and troughs which could be verified by the value of $\chi^2$ calculated in Table \ref{table1:chi2}. 
However, when performing a Bayesian analysis that considers variations in the effective number of neutrino species $N_\text{eff}$ across our three best inflationary models, the generalized Starobinsky inflationary model emerges as the favored model.

The article concludes by summarizing the key findings and comparing the performance of the three models in reproducing the observed CMB angular spectrum, highlighting their strengths and limitations.

\section{Acknowledgements}

We want to thank Professor Antony Lewis for his continuous support and assistance, always answering our questions on the CosmoCoffee Forum, and for his suggestion of using the \texttt{Cobaya} code. Additionally, we want to express our gratitude to Professor V\'ictor Miralles for the valuable discussions regarding the use of  \texttt{CAMB} code, and Professor Rafael Hernández-Jiménez for his valuable explanations about the use of \texttt{Cobaya} code. We also greatly appreciate the suggestions made by the anonymous referees, which have helped us improve our manuscript.
\bibliographystyle{unsrt}

\end{document}